\documentclass{emulateapj}

\shorttitle{Oxygen Chemistry in IRC+10216} \shortauthors{Ag\'undez
\& Cernicharo}

\begin{document}

\title{Oxygen Chemistry in the Circumstellar Envelope of the Carbon-Rich Star IRC+10216}

\author{Marcelino Ag\'undez and Jos\'e Cernicharo}
\affil{Departamento de Astrof\'isica Molecular e Infrarroja,
Instituto de Estructura de la Materia, CSIC, Serrano 121, E--28006
Madrid, Spain} \email{marce, cerni@damir.iem.csic.es}

\begin{abstract}

In this paper we study the oxygen chemistry in the C-rich
circumstellar shells of IRC+10216. The recent discoveries of
oxygen bearing species (water, hydroxyl radical and formaldehyde)
toward this source challenge our current understanding of the
chemistry in C-rich circumstellar envelopes. The presence of icy
comets surrounding the star or catalysis on iron grain surfaces
have been invoked to explain the presence of such unexpected
species. This detailed study aims at evaluating the chances of
producing O--bearing species in the C-rich circumstellar envelope
only by gas phase chemical reactions. For the inner hot envelope,
it is shown that although most of the oxygen is locked in CO near
the photosphere (as expected for a C/O ratio greater than 1), some
stellar radii far away species such as H$_2$O and CO$_2$ have
large abundances under the assumption of thermochemical
equilibrium. It is also shown how non-LTE chemistry makes very
difficult the CO$\rightarrow$H$_2$O, CO$_2$ transformation
predicted in LTE. Concerning the chemistry in the outer and colder
envelope, we show that formaldehyde can be formed through gas
phase reactions. However, in order to form water vapor it is
necessary to include a radiative association between atomic oxygen
and molecular hydrogen with a quite high rate constant. The
chemical models explain the presence of HCO$^+$ and predict the
existence of SO and H$_2$CS (which has been detected in a
$\lambda$ 3 mm line survey to be published). We have modeled the
line profiles of H$_2$CO, H$_2$O, HCO$^+$, SO and H$_2$CS using a
non-local radiative transfer model and the abundance profiles
predicted by our chemical model. The results have been compared to
the observations and discussed.

\end{abstract}


\section{Introduction}

IRC+10216 is a low mass AGB star losing mass at a rate of
2-4$\times$10$^{-5}$ M$_{\odot}$ yr$^{-1}$ in the form of a
molecular and dusty wind that produces an extended circumstellar
envelope (CSE). The processes of dredge-up that occurs during this
evolutionary late stage alter the elemental composition in the
stellar surface (that is roughly solar, C/O$<$1, during the main
sequence phase) resulting, in the case of IRC+10216, in a C/O
ratio greater than 1, in what is known as a carbon star.

The physical conditions in the vicinity of the photosphere of an
AGB star (typical temperatures of $\sim$2500 K and densities of
$\sim$10$^{14}$ cm$^{-3}$) make the material to be mainly
molecular, with a composition determined by local thermodynamic
equilibrium (LTE). It is now well established since the pioneering
work of \citet{tsu73} that the C/O ratio completely determines the
kind of chemistry taking place. The high stability of CO makes
this molecule to have a large abundance locking almost all of the
limiting reactant and allowing for the reactant in excess to form
either carbon bearing molecules when C/O$>$1, and the opposite,
i.e., oxygen bearing molecules, when C/O$<$1. This has been
extensively confirmed by observations. A look to the list of
molecules detected in IRC+10216 confirms that they are mostly
carbon bearing species (see table 1 in \citealt{cer00} and \S 3).
The only oxygen bearing molecule in C-rich AGB stars with a
significant abundance, apart from CO, is SiO \citep{mor75}.

However, some other O--bearing molecules such as H$_2$O
\citep{mel01,has06}, OH \citep{for03}, and H$_2$CO \citep{for04}
have recently been detected in IRC+10216. The existence of such
molecules has been interpreted as the result of evaporation of
cometary ices from a Kuiper belt analog in IRC+10216
\citep{for01}. The luminosity increase of the star, intrinsic to
the red giant phase, would have caused the ice sublimation and
subsequent release of water vapor to the gas phase in the
circumstellar shells. OH would be produced when water is
photodissociated by the interstellar UV field in the unshielded
outer envelope and H$_2$CO would be the photodissociation product
of an unknown parent molecule produced by sublimation of the ice
mantles of these comets.

An alternative explanation has been proposed by \citet{wil04} in
which H$_2$O would be produced through Fischer-Tropsch catalysis
on the surface of iron grains, that would be present in the
expanding envelope due to condensation from the gas phase of some
fraction of the available iron. Fischer-Tropsch catalysis breaks
the CO bond and produces H$_2$O and hydrocarbons such as CH$_4$.

O--bearing species have also been detected in the C-rich
protoplanetary nebula CRL618 \citep{her00}. However, the
production of these species has been interpreted by \citet{cer04}
as the result of a rich photochemistry in a region of high density
($\sim$10$^7$ cm$^{-3}$) and temperature (200-300 K) where large
complex carbon-rich molecules are also produced \citep{cer01}.

The aim of this paper is to investigate whether oxygen bearing
molecules could be produced in the carbon-rich expanding gas of
AGB stars by non-LTE mechanisms. LTE calculations provide a good
estimation of molecular abundances in the vicinity of the
photosphere but as the gas expands the temperature and density
decrease significantly and the chemical timescale increases,
making chemical kinetics dominant in determining the molecular
abundances. We describe the model of the circumstellar envelope in
\S 2. The reaction network is discussed in \S 3. The results of
the chemical model are presented in \S 4 together with comparisons
of radiative transfer calculations with available observations.
The conclusions are given in \S 5.

\section{The Circumstellar envelope}

We assume spherical symmetry for the circumstellar envelope. In
order to calculate the molecular abundances at different radii in
the CSE, we follow the history of a volume element of gas with a
given chemical composition travelling from the photosphere
($r$=$R_*$) to the end of the envelope ($r$$\sim$10$^{18}$ cm). We
build a system of differential equations which integration yields
the temporal evolution of the density of each gas species. The
time has to be interpreted as radial position provided the gas
travels outwards at a given velocity. Different processes are
considered during this travel depending on the position in the
CSE. Kinetic temperature (which determines the reaction rate
constants) and total gas density radial profiles are needed to
solve the system of equations. The values adopted for the
different parameters used to model the CSE are given in
table~\ref{tabl-param}.

For model purposes the CSE is considered to consist of three
different regions: the innermost region, the inner envelope, and
the intermediate and outer envelope.

\subsection{The innermost region}

\begin{figure}
\includegraphics[angle=0,scale=.44]{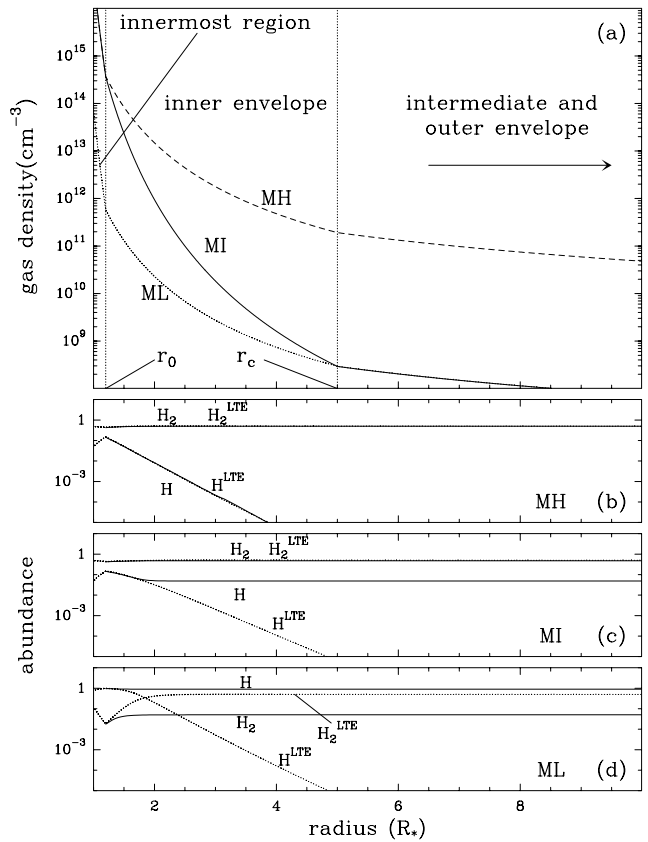}
\caption{(a) Three different density profiles considered for the
inner and innermost envelope: \textbf{--MH} [$K_n$=1,
$n(r_0)$=3.7$\times$10$^{14}$ cm$^{-3}$], \textbf{--MI}
[$K_n$=1.8565, $n(r_0)$=3.7$\times$10$^{14}$ cm$^{-3}$] and
\textbf{--ML} [$K_n$=1, $n(r_0)$=1.2 $\times$ 10$^{12}$
cm$^{-3}$]; all them use a r$^{-2}$ law for r$>$r$_c$. (b), (c)
and (d) show the abundances of H and H$_2$, relative to total
number of hydrogen nuclei [n(H)+2n(H$_2$)], as given by LTE
(dotted lines) and by chemical kinetics (solid lines) for the
three density profiles: (b)$\rightarrow$MH, (c)$\rightarrow$MI and
(c)$\rightarrow$ML. Note the strong dependence of the H/H$_2$
ratio on the densities considered.} \label{fig-fis-inn-vcte}
\end{figure}

This zone corresponds to the region between the photosphere and
$r_0$ (see Fig.~\ref{fig-fis-inn-vcte}a)
\notetoeditor{Fig.~\ref{fig-fis-inn-vcte} should appear with a 1
column width}. The adopted physical conditions are:

-- The \emph{temperature} is considered to vary as a power law of
the radius $r$
\begin{equation}
\displaystyle T(r) = T_* \times \Big(\frac{r}{R_*}\Big)^{-\alpha}
\end{equation}
where $T_*$ is the temperature at the photosphere and $R_*$ is the
star radius.

-- The \emph{density} profile is given by hydrostatic equilibrium.
Considering the above temperature law, the gas density can be
expressed as follows in terms of the value at $r_0$, $n(r_0)$
\begin{equation}
n(r) = n(r_{0}) \Big(\frac{r}{r_{0}}\Big)^{\alpha} exp
\bigg\{-\frac{G M_{*} m_H \mu}{k T_{*} R_{*}^{\alpha} (1-\alpha)}
\Big(r_{0}^{(\alpha-1)}-r^{(\alpha-1)}\Big) \bigg\}
\end{equation}
where $k$ is the Boltzmann constant, $G$ is the gravitational
constant, $M_*$ the mass of the star, $m_H$ is the mass of a
hydrogen atom and $\mu$ the mean molecular weight of gas.

The high densities and temperatures in this region allow to assume
that molecular abundances are given by a LTE calculation.

\subsection{The inner envelope}

The gas travels from $r_0$ to $r_c$ (see
Fig.~\ref{fig-fis-inn-vcte}a), where $r_c$ stands for condensation
radius. At this distance, grain formation is supposed to occur.
Radiation pressure on grains, together with momentum coupling of
gas and grains, make the gas accelerate up to a terminal expansion
velocity which remains constant beyond $r_c$. The mechanism
responsible for the transport of the gas from $r_0$ to $r_c$ is
somewhat controversial. Hydrodynamical models \citep{bow88} have
shown that pulsational driven shocks can gradually move the gas up
to $r_c$. \citet{wil98} have applied this approach for modelling
the chemistry in the inner envelope. An alternative mechanism
could be that the gas close to the photosphere pulsates around an
equilibrium position with an associated radial velocity very
similar to the escape velocity. This could lead to a scenario in
which shells of gas that are in levitation can eventually escape
from the surroundings of the photosphere and some of them could
reach $r_c$.

-- The \emph{temperature} profile in this region is given by
equation 1.

-- For the \emph{density} profile we follow the treatment of a
shocked extended region of a C-rich CSE of \citet{che92}
\begin{equation}
\displaystyle n(r) = n(r_0) exp \bigg\{-K_n \frac{G M_* m_H \mu
(1-\gamma^{2})}{k T_* R_*^{\alpha} (1-\alpha)}
\Big(r_0^{(\alpha-1)}-r^{(\alpha-1)}\Big) \bigg\}
\end{equation}
where $\gamma$ is defined in \citet{che92} and $K_n$ is explained
in the next paragraph.

The reference value $n(r_0)$ is not evident neither from
observations nor from models. Previous models of the inner
envelope of IRC+10216 \citep{che92,wil98} have used high values of
$n(r_0)$ based on hydrodynamical models for periodically shocked
Mira-like stars, which have density profiles as an output.
However, with such high $n(r_0)$ values, the extrapolation to
larger radii produces densities for the outer envelope much higher
than those obtained from the law of conservation of mass, in which
the absolute density is well determined by the mass loss rate and
the expansion velocity (see equations 5 and 6 in \S 2.3). To
reconcile the outer envelope density law with the high values in
the inner regions we introduce the arbitrary factor $K_n$ in
equation 3. We consider three different density laws for the inner
envelope, shown in Fig~\ref{fig-fis-inn-vcte}a, whose parameters
$n(r_0)$ and $K_n$ are given in table~\ref{tabl-param}: MI (I for
intermediate density); MH (H for high density), also considered
for comparing our results with the previous model of \citet{wil98}
although it overestimates the density for large radii; and finally
ML (L for low density). We point out that the chemical time scale
strongly depends on the density and also the H/H$_2$ ratio is
completely different in a low or high density situation (see
Figs.~\ref{fig-fis-inn-vcte}b, \ref{fig-fis-inn-vcte}c and
\ref{fig-fis-inn-vcte}d). We will comment this in \S 4.1.

Another issue is how the gas travels from $r_0$ to $r_c$. As
discussed above, different mechanisms could be operating. The
simplest assumption to model the chemistry is to consider that the
gas travels at constant velocity in this inner envelope. When
using this approach we will assume a velocity of 1 km/s.\\

-- \emph{SiO condensation onto SiC grains}. Considering non-LTE
effects, we investigated whether deposition of Si species (that
are specially refractory) on SiC grains could affect the oxygen
chemistry because SiO is the second more abundant O--bearing
molecule. The scheme for grain formation in IRC+10216 could be as
follows. The major type of grains are carbonaceous, most probably
amorphous carbon (A.C.) as indicated by fitting the IR spectrum
\citep{bag95}. These grains would originate in condensation
processes involving C$_2$H$_2$ (the most abundant C--bearing
molecule after CO) that would take place at $\sim$1000 K ($\sim$4
R$_*$). A minor type of grains would be SiC, as indicated by the
11.3 $\mu$m feature in the IR spectrum \citep{lor93}, that would
condense at $\sim$1500 K ($\sim$2 R$_*$). This condensation
sequence is due to an inverse greenhouse effect in both types of
grains (see \citealt{mcc82} for a theoretical approach and
\citealt{fre89} for experimental evidence). Although a more
correct approach would be a chemical mechanism leading to the
formation of grains from gaseous molecules, for our purposes we
describe the condensation of SiC grains (A.C. grains are not
considered because we are just interested in deposition of Si
species on SiC grains) as a two-step process: (1)
\emph{nucleation} from SiC gas molecules in the context of
homogeneous nucleation theory, which is basically described by the
nucleation rate $J_*$ (number of condensation nuclei formed per
unit time and unit volume); and (2) \emph{growth of grains} by
accretion of gas phase molecules, that is described by the grain
radius $a_{gr}$. The calculation of
these two magnitudes is detailed in Appendix A.\\

\begin{figure}
\includegraphics[angle=0,scale=.63]{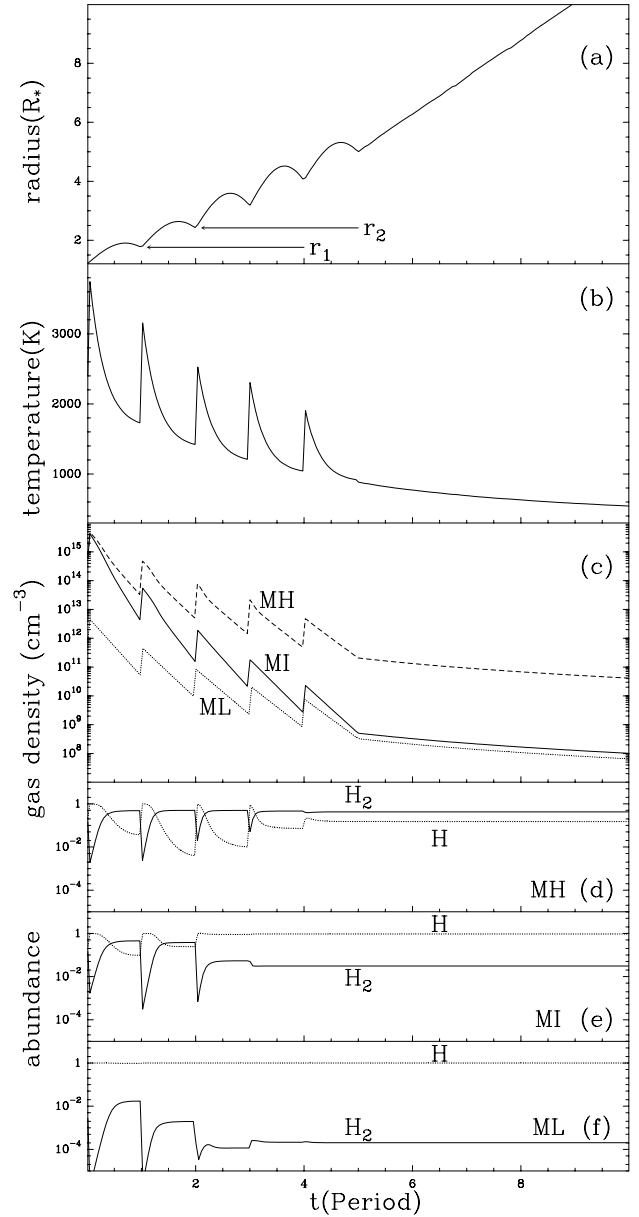}
\caption{Parameters for a volume element of gas that suffers a
history of 5 shocks: (a) trajectory; (b) temperature history; (c)
density histories for models MH, MI and ML. (d), (e) and (f) show
the H and H$_2$ abundances, relative to total number of hydrogen
nuclei, calculated by chemical kinetics for the three density laws
shown in (c).} \label{fig-fis-inn-shoc}
\end{figure}

\begin{deluxetable*}{rlr|llll|rlr}
\tabletypesize{\scriptsize} \tablecaption{Model
parameters\label{tabl-param}} \tablewidth{0pc} \startdata \hline
\hline
\multicolumn{3}{c}{}                                     & \multicolumn{4}{c}{} & \multicolumn{3}{c}{Abundances relative to H$_2$}\\
\multicolumn{3}{c}{}                                     & \multicolumn{4}{c}{} & \multicolumn{3}{c}{for parent species entering}\\
\multicolumn{3}{c}{IRC+10216 parameters}                 & \multicolumn{4}{c}{Neutral molecules considered} & \multicolumn{3}{c}{into the outer envelope}\\
\hline
\multicolumn{7}{c}{} \\
$[$1$]$ & $R_*$/cm                            & 6.5 10$^{13}$ & CH             & H$_2$          & N$_2$     & SiH        &           & H$_2$      & \\
$[$2$]$ & $M_*$/M$_{\odot}$                   & 2             & CH$_2$         & O$_2$          & NH        & SiH$_2$    & $[$18$]$  & H          & 4.5 10$^{-3}$ \\
$[$2$]$ & $r_0$/R$_*$                         & 1.2           & CH$_3$         & OH             & NH$_2$    & SiH$_3$    & $[$10$]$  & He         & 2.0 10$^{-1}$ \\
$[$1$]$ & $r_c$/$R_*$                         & 5             & CH$_4$         & H$_2$O         & NH$_3$    & SiH$_4$    & $[$7$]$   & CO         & 6.0 10$^{-4}$ \\
$[$1$]$ & $T_*$/K                             & 2320          & C$_2$          & HO$_2$         & CN        & Si$_2$     & $[$19$]$  & C$_2$H$_2$ & 8.0 10$^{-5}$ \\
$[$3$]$ & $\alpha$                            & 0.6           & C$_2$H         & H$_2$O$_2$     & HCN       & SiC        & $[$19$]$  & HCN        & 4.9 10$^{-5}$ \\
$[$2$]$ & $n(r_0)^{MH,MI}$/cm$^{-3}$          & 3.7 10$^{14}$ & C$_2$H$_2$     & CO             & NS        & SiCH       & $[$10$]$  & N$_2$      & 8.0 10$^{-5}$ \\
        & $n(r_0)^{ML}$/cm$^{-3}$             & 1.2 10$^{12}$ & C$_2$H$_3$     & CO$_2$         & SH        & SiCH$_2$   & $[$13$]$  & C$_2$H$_4$ & 2.0 10$^{-8}$ \\
        & $K_n^{MH,ML}$                       & 1             & C$_2$H$_4$     & HCO            & H$_2$S    & SiC$_2$    & $[$11$]$  & CH$_4$     & 3.5 10$^{-6}$ \\
        & $K_n^{MI}$                          & 1.8565        & C$_2$H$_5$     & H$_2$CO        & CS        & Si$_2$C    & $[$11$]$  & NH$_3$     & 1.7 10$^{-7}$ \\
$[$4$]$ & $P$/days                            & 650           & C$_2$H$_6$     & CH$_3$O        & HCS       & SiN        & $[$11$]$  & SiH$_4$    & 2.2 10$^{-7}$ \\
$[$3$]$ & $\gamma$                            & 0.89          & C$_3$          & CH$_2$OH       & H$_2$CS   & SiNH       & $[$14$]$  & H$_2$S     & 6.0 10$^{-10}$ \\
$[$3$]$ & $\mu$(r$<$r$_c$)/amu                & 1.7           & C$_3$H         & CH$_3$OH       & S$_2$     & SiO        & $[$15$]$  & CS         & 1.2 10$^{-7}$ \\
$[$5$]$ & $v_{exp}$/(km/s)                    & 14.5          & C$_3$H$_2$     & HCCO           & CS$_2$    & SiO$_2$    & $[$16$]$  & SiS        & 4.3 10$^{-6}$ \\
$[$6$]$ & $\dot{M}$/(M$_{\odot}$ yr$^{-1}$)   & 3 10$^{-5}$   & C$_4$          & NO             & SO        & SiS        & $[$17$]$  & SiC$_2$    & 5.0 10$^{-8}$ \\
$[$7$]$ & distance/pc                         & 150           & C$_4$H         & NCO            & SO$_2$    &            & $[$11$]$  & SiO        & 8.0 10$^{-7}$ \\
$[$8$]$ & $\tau_{1000}(r=10^{16})$            & 12.7          & C$_4$H$_2$     &                & OCS       &            &           &            &     \\
\enddata
\tablerefs{ $[$1$]$ \citealt{rid88}; $[$2$]$ \citealt{wil98};
$[$3$]$ \citealt{che92}; $[$4$]$ \citealt{wit80}; $[$5$]$
\citealt{cer00}; $[$6$]$ \citealt{gla96}; $[$7$]$ \citealt{cro97};
$[$8$]$ \citealt{dot98}; $[$9$]$ \citealt{rou91}; $[$10$]$
calculated according to LTE, solar elemental abundances and a C/O
ratio of 2; $[$11$]$ \citealt{kea93}; $[$12$]$ \citealt{wie91};
$[$13$]$ \citealt{gol87}; $[$14$]$ Abundance derived from the
1$_{1,0}$-1$_{0,1}$ line of ortho-H$_2$S observed in
\citealt{cer00} in the optically thin and LTE limit assuming
T$_{rot}$ = 20 K and considering that H$_2$S is present for r$>$50
R$_*$; $[$15$]$ \citealt{hen85}; $[$16$]$ \citealt{boy94};
$[$17$]$ upper limit by \citealt{gen95}; $[$18$]$ upper limit by
\citealt{bow87}; $[$19$]$ \citealt{fon06}.}
\end{deluxetable*}

-- \emph{Shocks}. An alternative to the mechanism of the gas
travelling at constant velocity in the inner envelope is to
consider the effect of gas driven by shocks. Molecular abundances
are considerably affected by the steep changes in temperature and
density associated to the shocks. We follow the approach of
\citet{wil98} but in a more simplistic way. The history of a
volume element of gas to reach $r_c$ from $r_0$ after a certain
number of shocks is represented in Figs.~\ref{fig-fis-inn-shoc}a,
\ref{fig-fis-inn-shoc}b and \ref{fig-fis-inn-shoc}c, and is
described below \notetoeditor{Fig.~\ref{fig-fis-inn-shoc} should
appear with a 1 column width}. With one shock and subsequent
relaxation, the gas moves from a position $r_1$ to $r_{2}$
($r_{2}$$>$$r_{1}$) following a harmonic trajectory. The gas in
$r_1$ with temperature $T_1$ and density $n_1$ suffers a shock and
we assume that the immediately post-shocked gas increases its
temperature by a factor 2 and its density by a factor 10 (typical
values obtained by \citealt{wil98} from hydrodynamical
considerations). Then the gas relaxes its temperature and density
exponentially until reaching the new radius $r_2$ (with
temperature $T_2$ and density $n_2$) in a time interval equal to
the period of pulsation of the star $P$. In this way the gas
progressively moves outwards until reaching $r_c$, beyond which an
expansion at constant velocity $v_{exp}$ takes place.

In Fig.~\ref{fig-fis-inn-shoc} we plot the trajectory, temperature
and the three density laws introduced previously. Note that again
the H/H$_2$ ratio strongly depends on the total gas density (see
Figs.~\ref{fig-fis-inn-shoc}d, \ref{fig-fis-inn-shoc}e and
\ref{fig-fis-inn-shoc}f).

\subsection{The intermediate and outer envelope}

It extends from $r_c$. The gas expands adiabatically at constant
velocity $v_{exp}$. The molecules in the very outer envelope are
no longer shielded by the dust of the CSE against the interstellar
UV field, and are then photodissociated. Cosmic rays also play a
role ionizing some species.

-- The \emph{temperature} profile is taken from the fit of
\citet{mam88} to the results of \citet{kwa82} for the outer
envelope of IRC+10216
\begin{equation}
\displaystyle T(r) = max\Big[10, 14.6 \Big(\frac{r}{9 \times
10^{16}}\Big)^{-p}\Big]
\end{equation}
where $r$ is expressed in cm and $p$ takes values of 0.72 for
$r$$<$9$\times$10$^{16}$ and 0.54 for $r$$>$9$\times$10$^{16}$.

-- The \emph{density} profile is given by the law of conservation
of mass:
\begin{equation}
\displaystyle \dot{M} = n(r) m_H \mu v_{exp} 4 \pi r^2
\end{equation}
which, with the values given in table~\ref{tabl-param} and
assuming that hydrogen is mostly molecular ($\mu$$\sim$2 amu),
results in
\begin{equation}
n(r)=\frac{3.1 \times 10^{37}}{r(cm)^2} \quad cm^{-3}
\end{equation}

-- The photodissociation rate $\Gamma_i$ for a molecule $i$ at a
radius $r$ depends on the UV field at that radius:
\begin{equation}
\Gamma_i(r) = \int_{91.2}^{\lambda_i} 4\pi J_{\lambda}(r)
\sigma_i(\lambda) d\lambda
\end{equation}
where $4\pi J_{\lambda}(r)$ is the UV field at $r$ in photons
cm$^{-2}$ s$^{-1}$ nm$^{-1}$, $\sigma_i(\lambda)$ is the
photodissociation cross section and the integral extends from 91.2
nm (Lyman cutoff) to a threshold value $\lambda_i$ that depends on
each molecule. The calculation of each $\Gamma_i(r)$ involves the
knowledge of the cross section and the solution of the UV
radiative transfer for calculating $J_{\lambda}$ at each radius
$r$, which will depend in a first approximation on the amount of
dust that surrounds the point $r$. In practice, photodissociation
rates are usually expressed in the literature \citep{let00} for
plane-parallel geometry as a function of the visual extinction
$A_v$ (in magnitudes) measured along the direction normal to the
infinite plane
\begin{equation}
\Gamma_{i}(A_v) = A_i exp(-C_i A_v)
\end{equation}
with specific parameters $A_i$ and $C_i$ for each molecule. For
our purposes we have adopted this simple approach just correcting
for the geometrical difference from plane-parallel to spherical.
The procedure used for such correction as well as the resulting
$A_v$ radial profile are detailed in appendix B.

\section{The Chemistry}

We consider solar elemental abundances (oxygen from
\citealt{all01} and the rest from \citealt{cox00}) and assume a
C/O ratio of 2. The chemistry in IRC+10216 is dominated by carbon
bearing molecules. For carbon chain radicals C$_n$H see
\citet{tuc74,tha85,gue78,cer86,gue87a,gue97,cer96c8h}. For carbon
chain radicals C$_n$N see \citet{wil71,gue77,gue98}. For
cyanopolyynes HC$_{2n+1}$N see \citet{mor71,mor76,win78,bel82}.
For sulfur--carbon chain molecules C$_n$S see
\citet{wil71,cer87,bel93}. And for silicon bearing molecules
SiC$_n$ see \citet{cer89,tha84,app99,ohi89}. However, since we are
just interested in oxygen chemistry, in our model we don't include
very long carbon chains but only the 65 neutral molecules given in
table~\ref{tabl-param}.

As initial abundances for the inner envelope we take those given
by LTE at $r_0$, while for the outer shells we adopt initial
abundances for some parent molecules taken from the literature or
from a LTE calculation when no data are available (see
table~\ref{tabl-param}). This procedure is motivated by the
processes occurring in the intermediate envelope (specially on
grain surfaces) that presumably alter the molecular abundances
(hydrides such as CH$_4$, NH$_3$, SiH$_4$ and H$_2$S are supposed
to form via such processes).

\begin{deluxetable}{l@{ }l@{ }l@{ }r@{ }r}
\tabletypesize{\footnotesize} \tablecaption{Reactions relevant for
oxygen chemistry\label{tabl-reac}} \tablewidth{0pc} \startdata
\hline \hline
 & Reaction & A & n & C \\
\hline
\multicolumn{1}{c}{} & \multicolumn{4}{c}{Important reactions in the INNER envelope} \\
R1  & H + H + H$_2$ $\rightarrow$ H$_2$ + H$_2$                          & 8.6$\times$10$^{-33}$    & -0.60 & 0 \\
R2  & H + H + H $\rightarrow$ H$_2$ + H                                  & 8.8$\times$10$^{-33}$    & 0.00  & 0 \\
R3  & H + H + He $\rightarrow$ H$_2$ + He                                & 6.1$\times$10$^{-33}$    & -0.13 & -39 \\
R4  & H$_2$ + H$_2$ $\rightarrow$ H + H + H$_2$                          & 8.4$\times$10$^{-09}$    & -0.24 & 52043 \\
R5  & H$_2$ + H $\rightarrow$ H + H + H                                  & 8.6$\times$10$^{-09}$    & 0.36  & 52043 \\
R6  & H$_2$ + He $\rightarrow$ H + H + He                                & 5.9$\times$10$^{-09}$    & 0.23  & 52003 \\
R7  & Si + CO $\rightarrow$ SiO + C                                      & 1.3$\times$10$^{-09}$    & 0.00  & 34513 \\
R8  & SiO + C $\rightarrow$ Si + CO                                      & 1.0$\times$10$^{-09}$    & -0.23 & 1291 \\
R9  & C + H$_2$ $\rightarrow$ CH + H                                     & 3.1$\times$10$^{-10}$    & 0.16  & 11894 \\
R10 & OH + Si $\rightarrow$ SiO + H                                      & 1.0$\times$10$^{-10}$    & 0.00  & 0 \\
R11 & OH + H$_2$ $\rightarrow$ H$_2$O + H                                & 2.2$\times$10$^{-12}$    & 1.43  & 1751 \\
R12 & SiO + CO $\rightarrow$ CO$_2$ + Si                                 & 4.2$\times$10$^{-13}$    & 0.67  & 32225 \\
R13 & CO$_2$ + H$_2$ $\rightarrow$ H$_2$O + CO                           & 3.2$\times$10$^{-7}$     & 1.53  & 56906 \\
R14 & OH + CO $\rightarrow$ CO$_2$ + H                                   & 1.2$\times$10$^{-13}$    & 0.95  & -73 \tablenotemark{a} \\
R15 & O + H$_2$ $\rightarrow$ OH + H                                     & 3.5$\times$10$^{-13}$    & 2.60  & 3241 \\
R16 & CO$_2$ + Si $\rightarrow$ SiO + CO                                 & 2.7$\times$10$^{-11}$    & 0.00  & 282 \\
\multicolumn{1}{c}{} & \multicolumn{4}{c}{Important reactions in the OUTER envelope} \\
R17 & O + NH$_2$ $\rightarrow$ OH + NH                                   & 1.2$\times$10$^{-11}$    & 0.00  & 0 \\
R18 & O + CH$_3$ $\rightarrow$ H$_2$CO + H                               & 1.4$\times$10$^{-10}$    & 0.00  & 0 \\
R19 & OH + S $\rightarrow$ SO + H                                        & 6.6$\times$10$^{-11}$    & 0.00  & 0 \\
R20 & O + SH $\rightarrow$ SO + H                                        & 1.2$\times$10$^{-10}$    & 0.00  & -74 \tablenotemark{c} \\
R21 & O + NH $\rightarrow$ NO + H                                        & 1.2$\times$10$^{-10}$    & 0.00  & 0 \\
R22 & CN + OH $\rightarrow$ NCO + H                                      & 1.4$\times$10$^{-10}$    & 0.00  & 0 \\
R23 & S + CH$_3$ $\rightarrow$ H$_2$CS + H                               & 1.4$\times$10$^{-10}$    & 0.00  & 0 \\
R24 & O + OH $\rightarrow$ O$_2$ + H                                     & 1.8$\times$10$^{-11}$    & 0.00  & -175 \tablenotemark{b} \\
R25 & O + CH$_2$ $\rightarrow$ HCO + H                                   & 5.0$\times$10$^{-11}$    & 0.00  & 0 \\
R26 & $^{13}$C$^+$ + $^{12}$CO $\rightarrow$ $^{12}$C$^+$ + $^{13}$CO    & 3.4$\times$10$^{-10}$    & -0.50 & 1.3 \\
R27 & SiO + C$^+$ $\rightarrow$ Si$^+$ + CO                              & 5.4$\times$10$^{-10}$    & 0.00  & 0 \\
R28 & H$_2^+$ + H$_2$ $\rightarrow$ H$_3^+$ + H                          & 2.1$\times$10$^{-9}$     & 0.00  & 0 \\
R29 & H$_3^+$ + CO $\rightarrow$ HCO$^+$ + H$_2$                         & 1.7$\times$10$^{-9}$     & 0.00  & 0 \\
R30 & HCO$^+$ + HCN $\rightarrow$ CO + HCNH$^+$                          & 3.1$\times$10$^{-9}$     & 0.00  & 0 \\
R31 & HCO$^+$ + C$_2$H$_2$ $\rightarrow$ CO + C$_2$H$_3^+$               & 1.4$\times$10$^{-9}$     & 0.00  & 0 \\
R32 & SiS + C$^+$ $\rightarrow$ SiS$^+$ + C                              & 2.3$\times$10$^{-9}$     & 0.00  & 0 \\
R33 & SiS$^+$ + H $\rightarrow$ SH + Si$^+$                              & 1.9$\times$10$^{-9}$     & 0.00  & 0 \\
R34 & SiS + S$^+$ $\rightarrow$ SiS$^+$ + S                              & 3.2$\times$10$^{-9}$     & 0.00  & 0 \\
R35 & SiO$^+$ + CO $\rightarrow$ CO$_2$ + Si$^+$                         & 7.9$\times$10$^{-10}$    & 0.00  & 0 \\
R36 & O + H $\rightarrow$ OH + h$\nu$                                    & 9.9$\times$10$^{-19}$    & -0.38 & 0 \\
R37 & S + CO $\rightarrow$ OCS + h$\nu$                                  & 1.6$\times$10$^{-17}$    & -1.50 & 0 \\
R38 & HCO$^+$ + e$^-$ $\rightarrow$ CO + H                               & 1.1$\times$10$^{-7}$     & -1.00 & 0 \\
R39 & H$_3$CS$^+$ + e$^-$ $\rightarrow$ H$_2$CS + H                      & 3.0$\times$10$^{-7}$     & -0.50 & 0 \\
R40 & H$_2$CO$^+$ + e$^-$ $\rightarrow$ HCO + H                          & 1.0$\times$10$^{-7}$     & -0.50 & 0 \\
R41 & SiO + h$\nu$ $\rightarrow$ Si + O                                  & 1.0$\times$10$^{-10}$    & 0.00  & 2.3 \\
R42 & H$_2$O + h$\nu$ $\rightarrow$ OH + H                               & 5.9$\times$10$^{-10}$    & 0.00  & 1.7 \\
R43 & CH$_4$ + h$\nu$ $\rightarrow$ CH$_3$ + H                           & 2.2$\times$10$^{-10}$    & 0.00  & 2.2 \\
R44 & H$_2$ + CR $\rightarrow$ H$_2^+$ + e$^-$                           & 1.2$\times$10$^{-17}$    & 0.00  & 0 \\
\multicolumn{1}{c}{} & \multicolumn{4}{c}{Reactions not considered in previous chemical models \tablenotemark{d}} \\
R45 & O + H$_2$ $\rightarrow$ H$_2$O + h$\nu$                            & 1.0$\times$10$^{-18}$    & 0.00  & 0 \\
R46 & HCO$^+$ + H$_2$ $\rightarrow$ H$_3$CO$^+$ + h$\nu$                 & 5.0$\times$10$^{-15}$    & 0.00  & 0 \\
\enddata
\tablecomments{Rate constants for termolecular and bimolecular
reactions are given by k=A$\times$(T/300)$^n$$\times$exp(-C/T),
for photodissociations (h$\nu$) are given by k=A$\times$exp(-C
A$_v$) while for reactions with cosmic rays (CR) are given by k=A.
Units are $cm^{6} s^{-1}$ for termolecular reactions, $cm^{3}
s^{-1}$ for bimolecular reactions and $s^{-1}$ for
photodissociations and reactions with cosmic rays.}
\tablenotetext{a}{k(T$<$80K)=k(T=80K)}
\tablenotetext{b}{k(T$<$167K)=k(T=167K)}
\tablenotetext{c}{k(T$<$298K)=k(T=298K)} \tablenotetext{d}{See
text for a discussion on these rate constants.}
\end{deluxetable}

\begin{deluxetable*}{llrrrrrrr}
\tabletypesize{\scriptsize} \tablecaption{The first 5 reactions of
the full chemical kinetics mechanism\label{tabl-allreac}}
\tablewidth{0pc} \startdata \hline
N & Reaction & A & n & C & T$_1$ & T$_2$ & T$_{range}$ & Ref \\
\hline
1 & H + H + H2 $<=>$ H2 + H2   &  8.640E-33 & -0.60 &     0.00 &   50. & 5000. & ALL$_{--}$T & NIST \\
2 & H + H + He $<=>$ H2 + He   &  6.070E-33 & -0.13 &   -39.00 &   77. & 2000. & ALL$_{--}$T & NIST \\
3 & H + H + H $<=>$ H2 + H     &  8.820E-33 &  0.00 &     0.00 &   50. & 5000. & ALL$_{--}$T & NIST \\
4 & CH + He $<=>$ C + H + He   &  3.160E-10 &  0.00 & 33704.00 & 2500. & 3800. & LOW$_{--}$T & NIST \\
5 & CH + H2 $<=>$ C + H + H2   &  8.830E-10 &  0.00 & 33704.00 & 2500. & 3800. & LOW$_{--}$T & NIST \\
\enddata
\tablecomments{The full version of table~\ref{tabl-allreac} is
available in the electronic edition of the {\it Astrophysical
Journal}.}
\end{deluxetable*}

Different types of reactions dominate the chemistry depending on
the region of the CSE considered. The temperatures in the inner
envelope are not high enough to make ions abundant but allow for
reactions with activations energies of up to a few tens of
thousands of K to occur. Therefore, important processes are
termolecular reactions, its reverse (thermal dissociation) and
bimolecular reactions between neutrals. The outer envelope
chemistry is dominated by photodissociations, radical-molecule
reactions without activation energy and radiative associations.
Although all the species detected in IRC+10216, except HCO$^+$,
are neutral, ionic chemistry is necessary to explain the formation
of some species. Therefore, when modelling the chemistry in the
outer envelope we add 80 ionic species (mainly the positive ions
and protonated species of the neutral molecules of
table~\ref{tabl-param}) and include the subsequent reactions in
which they are involved. The rate constants have been taken from
different sources: databases such as NIST Chemical Kinetics
Database\footnote{The NIST Chemical Kinetics Database is available
on the World Wide Web at http://kinetics.nist.gov/index.php.},
UMIST Database 1999 \citep{let00} and osu.2003 Database
\citep{smi04}; estimations for reactions involving S-- and
Si--bearing species from \citet{wil98}; combustion mechanisms such
as GRI--Mech \citep{smi99} or the one of A. Konnov \citep{kon00};
and from a revision of the last published data of rate constants
for the type of reactions mentioned above. When no data are
available for the whole temperature range studied here (10-4000 K)
we have either extrapolated the expression or fixed the rate
constant to its value at the nearest temperature for which it is
known. When rate constants for reverse reactions were unknown or
uncertain, detailed balance has been applied for calculating them
from thermochemical properties of the species involved.
Photodissociation rates have been taken from the quoted databases
when available, or assumed to be equal to those of similar
molecules otherwise to ensure that all molecules are dissociated
at some radius within the envelope. CO and H$_2$ are known to be
affected by self-shielding against interstellar UV photons due to
their large abundance. CO photodisssociation rate as a function of
radial position is taken from \citet{dot98}, who studied the CO
case specifically for IRC+10216. For H$_2$ we use the same result
correcting for the different unattenuated photodissociation rate.

The most important reactions for oxygen chemistry are shown in
table~\ref{tabl-reac} \notetoeditor{ If possible
table~\ref{tabl-reac} should appear with a 1 column width}.
Table~\ref{tabl-allreac} contains the complete set of reactions.

\section{Results and discussion}

\subsection{Oxygen chemistry in the inner envelope}

In this section we (i) first compare the results of chemical
kinetics (assuming an expansion at constant velocity of 1 km
s$^{-1}$) with those of chemical equilibrium, (ii) secondly we
discuss the effects that SiO depletion from the gas phase could
have on oxygen chemistry and (iii) finally we consider the
chemistry with shocks.

(i) The densities considered in the inner envelope affect, among
other parameters, the H/H$_2$ ratio.
Figs.~\ref{fig-fis-inn-vcte}b, \ref{fig-fis-inn-vcte}c and
\ref{fig-fis-inn-vcte}d show the H$_2$ and H abundances for models
MH, MI and ML. The high densities of model MH make the chemical
time scale to be lower than the dynamical time scale associated to
an expansion at 1 km/s; therefore, reactions R1 to R6
\begin{displaymath}
\begin{array}{llcl}
R1, R2, R3 \qquad & H + H + M & \rightarrow & H_2 + M \\
R4, R5, R6 \qquad &  H_2 + M & \rightarrow & H + H + M \\
\end{array}
\end{displaymath}
are rapid enough to produce LTE abundances for H and H$_2$. The
steep decay in density of model MI makes that at $\sim$2 R$_*$
reactions R1 to R6 become too slow, so that the H abundance is
frozen at a value of 5$\times$10$^{-2}$. Both models MH and MI
allow hydrogen to enter the intermediate shells and outer envelope
mostly as H$_2$. Model MI has a low photospheric density and
therefore hydrogen exists in atomic form rather than molecular.
For larger radii the decrease in temperature makes molecular the
most stable form. However, chemical kinetics does not allow for a
H$\rightarrow$H$_2$ transformation and hydrogen enters the outer
envelope in atomic form. It is seen how both the density at the
photosphere and the density radial profile affect the H/H$_2$
ratio. There are observational constraints about this ratio that
suggest that hydrogen is mainly molecular in the CSE (based on an
upper limit for the H$_2$ mass in the CSE and on an estimation of
the H mass from observations of the 21 cm line, see discussion in
\citealt{gla96}). This suggests a high density scenario, such as
the one of models MH or MI, for the inner envelope.

\begin{figure*}
\includegraphics[angle=-90,scale=.72]{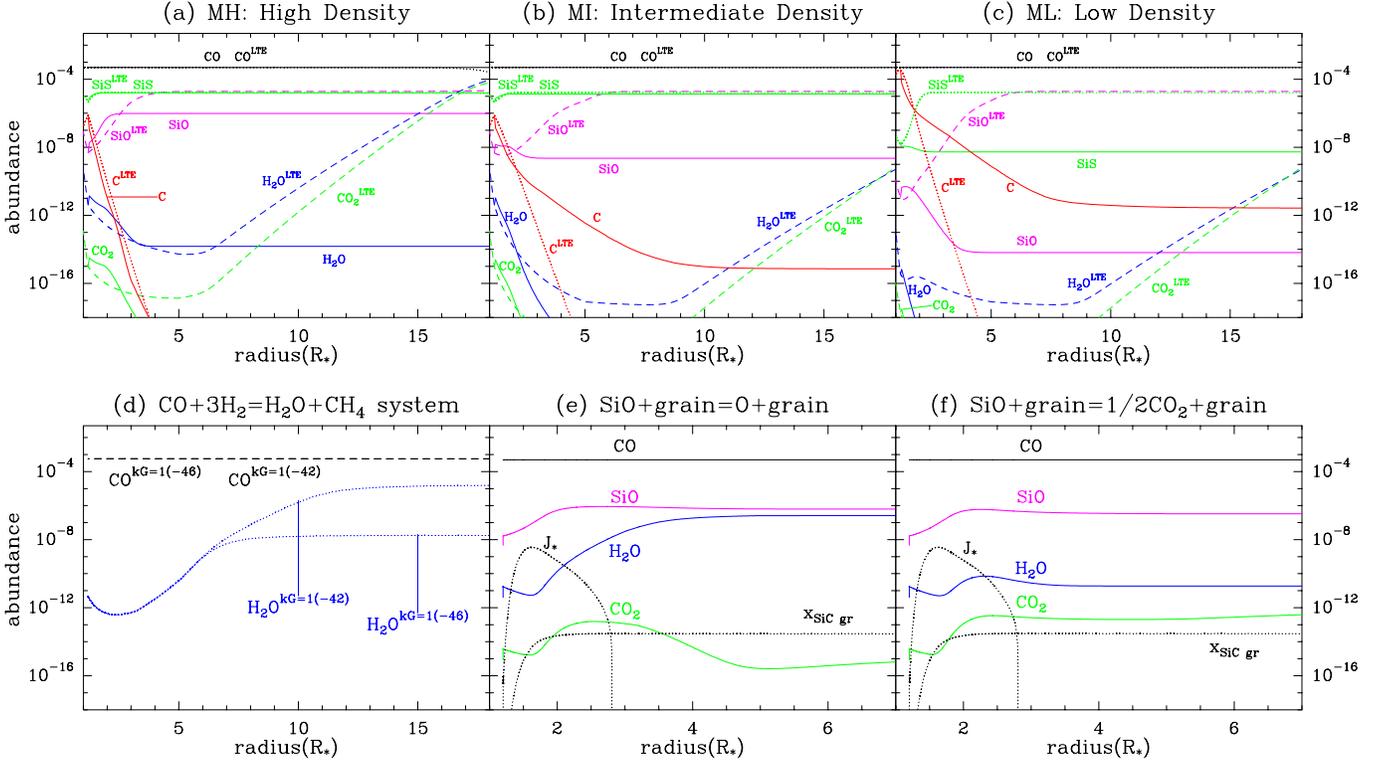}
\caption{Abundances, relative to total number of hydrogen nuclei,
of some oxygen bearing molecules. $-$(a), (b) and (c): abundances
given by LTE (dotted and dashed lines) and by chemical kinetics
(solid lines) assuming a constant velocity expansion of 1 km/s and
the density profiles MH$\rightarrow$(a), MI$\rightarrow$(b) and
ML$\rightarrow$(c). $-$(d): abundances of CO and H$_2$O for a
CO,H$_2$,CH$_4$,H$_2$O system with $k_G$ (see text) equal to
10$^{-42}$ and 10$^{-46}$ cm$^{9}$ s$^{-1}$. $-$(e) and (f):
abundances given by chemical kinetics with the density profile MH
when we consider that 90\% of SiO gas is deposited onto SiC grains
and that the corresponding oxygen is released to the gas phase
either as atomic oxygen$\rightarrow$(e) or as
CO$_2$$\rightarrow$(f). The nucleation rate of SiC grains $J_*$
(with units of nuclei cm$^{-3}$ s$^{-1}$) and its abundance
$x_{SiC gr}$, relative to total number of hydrogen nuclei, are
also plotted as dotted lines. See the electronic edition of the
Journal for a color version of this figure}
\label{fig-quim-inn-vcte}
\end{figure*}

In what concerns oxygen chemistry, according to chemical
equilibrium, for the densities and temperatures prevailing in the
inner envelope, almost all the oxygen available in a C-rich gas is
locked in CO, with SiO being 2-5 orders of magnitude less
abundant. Any other O--bearing molecule has an LTE abundance
always below $\sim$10$^{-10}$. But some stellar radii away from
the photosphere, where the temperature has decreased below
$\sim$700 K, the chemical system experiences a significant change
and molecules such as H$_2$O and CO$_2$ become very abundant in
LTE (see dotted lines in Figs.~\ref{fig-quim-inn-vcte}a,
\ref{fig-quim-inn-vcte}b and \ref{fig-quim-inn-vcte}c
\notetoeditor{Fig.~\ref{fig-quim-inn-vcte} should appear with a 2
column width}). This is related to the fact that the most stable
form for both carbon and oxygen is the CO molecule in a high
temperature regime, whereas for a low temperature regime carbon is
preferably locked in CH$_4$ and oxygen in H$_2$O \citep{tsu00}, as
is easily seen from the analysis of the equilibrium constant of
this reaction:
\begin{equation}
CO + 3 H_2 \rightleftharpoons CH_4 + H_2O
\end{equation}
which can be fitted by the expression
\begin{equation}
\displaystyle K_{eq}(T) = 1.58 \times 10^{-50}
\Big(\frac{T}{300}\Big)^{0.09} exp(24427/T)
\end{equation}
for a temperature interval of 200-2000 K ($K_{eq}$ with units of
(molecules/cm$^{3}$)$^{-2}$). The formation of CH$_4$ and H$_2$O
from CO and H$_2$ is exothermic and involves less particles, so
according to Le Chatelier's principle the CH$_4$/H$_2$O mixture
dominates over the CO/H$_2$ mixture at low temperatures and high
densities. In our case, as temperature decreases so does density.
Thus, a total CO$\rightarrow$H$_2$O transformation is only
produced for T$\lesssim$400 K (r$\gtrsim$ 18 R$_*$) in model MH
(see Fig.~\ref{fig-quim-inn-vcte}a) and for larger radii in models
MI and ML. Despite a total CO$\rightarrow$H$_2$O transformation
takes place beyond $\sim$20 R$_*$, the three models show a
considerably increase in the LTE abundance of H$_2$O beyond
$\sim$10 R$_{*}$.

In order to have LTE abundances in the inner envelope, the
chemical time scale $\tau_{chem}$ has to be shorter than the
dynamical time scale $\tau_{dyn}$ associated to the expansion for
the gas being able to readapt its molecular abundances to the
actual physical conditions (temperature and density) as it
expands. The situation in the inner envelope of IRC+10216 is such
that $\tau_{chem}$$<$$\tau_{dyn}$ in the vicinity of the
photosphere, but as temperature and density fall with radial
distance, chemical reactions become slower and for a certain radio
(that depends on the chemical species)
$\tau_{chem}$$>$$\tau_{dyn}$. There, molecular abundances "freeze"
and the LTE scenario disappears.

The LTE abundance of CO remains constant from the photosphere to
$\sim$20 R$_*$. Thus, whether chemical kinetics is rapid (LTE is
reproduced) or slow ("freezing" effect) the CO abundance won't be
modified. The situation is different for SiO (see
Figs.~\ref{fig-quim-inn-vcte}a, \ref{fig-quim-inn-vcte}b and
\ref{fig-quim-inn-vcte}c). Its LTE abundance at $r_{0}$ is
$\sim$10$^{-8}$ while at $\sim$3-5 R$_{*}$ it increases to
$\sim$10$^{-5}$. For this abundance enhancement to occur as the
gas expands, SiO has to take the oxygen from CO via reaction R7
\begin{displaymath}
\begin{array}{llcl}
R7 \qquad & Si + CO & \rightarrow & SiO + C \\
\end{array}
\end{displaymath}
which has a low rate and the above considerations about
$\tau_{chem}$ and $\tau_{dyn}$ apply. Only high densities and/or
low expansion velocities will make SiO reach this abundance of
$\sim$10$^{-5}$. For the three density models considered and the
expansion velocity of 1 km/s assumed, only the MH model makes SiO
reach an abundance near its LTE value, whereas MI and ML lead to a
non-LTE situation in which SiO abundance remains low. The
abundance of atomic C has a great impact on SiO. In models MI and
ML it remains well above its LTE value while in MH it follows the
decreasing LTE profile being destroyed by reaction R9
\begin{displaymath}
\begin{array}{llcl}
R9 \qquad & C + H_2 & \rightarrow & CH + H \\
\end{array}
\end{displaymath}
The H "freezing" effect of models MI and ML makes the reverse of
reaction R9 to proceed at a higher rate that in a LTE situation
making C overabundant which, through reaction R8
\begin{displaymath}
\begin{array}{llcl}
R8 \qquad & SiO + C & \rightarrow & Si + CO \\
\end{array}
\end{displaymath}
destroys SiO making it underabundant with respect to a LTE
situation. Again a high density scenario is required for
explaining the SiO observational abundance of 8$\times$10$^{-7}$
(estimated by \citealt{kea93} through IR ro-vibrational lines)
that could be due either to this "freezing" effect or to depletion
on SiC grains after reaching its LTE abundance of $\sim$10$^{-5}$.
Neither of the models run made any other O--bearing molecule to
have a significant abundance. In this high temperature regime
H$_2$O is related to SiO by two routes: (1) competition for OH via
reactions R10 and R11
\begin{displaymath}
\begin{array}{llcl}
R10 \qquad & OH + Si & \rightarrow & SiO + H \\
R11 \qquad & OH + H_2 & \rightarrow & H_2O + H \\
\end{array}
\end{displaymath}
and (2) through CO$_2$, directly via reactions R12+R13
\begin{displaymath}
\begin{array}{llcl}
R12 \qquad & SiO + CO & \rightarrow & CO_2 + Si \\
R13 \qquad & CO_2 + H_2 & \rightarrow & H_2O + CO \\
\end{array}
\end{displaymath}
or involving OH which is converted into CO$_2$ via reaction R14
\begin{displaymath}
\begin{array}{llcl}
R14 \qquad & OH + CO & \rightarrow & CO_2 + H \\
\end{array}
\end{displaymath}
The following scheme shows these relations:
\begin{displaymath}
\begin{array}{ccccc}
\displaystyle{SiO} & \displaystyle{\stackrel{R10}{\longleftarrow}} & \displaystyle{OH} & \displaystyle{\stackrel{R11}{\longrightarrow}} & \displaystyle{H_2O} \\
 & \scriptstyle{R12} \displaystyle{\searrow} & \displaystyle{\downarrow} \scriptstyle{R14} & \displaystyle{\nearrow} \scriptstyle{R13} & \\
 & & \displaystyle{CO_2} & & \\
\end{array}
\end{displaymath}
where the connection between the species is produced by the
reactions shown above and also by their reverse processes, so we
have a reversible system. Nevertheless, these mechanisms are not
efficient in producing H$_2$O from SiO. For H$_2$O to take the
oxygen from CO the chances are scarce because when the LTE
abundance of H$_2$O increases, the temperatures ($<$700 K) make
that the possible chemical routes for a CO$\rightarrow$H$_2$O
conversion will be too slow due to the large activation energy for
breaking the CO bond. At this point, we would like to know if we
miss some important reactions, rapid enough for allowing this
CO$\rightarrow$H$_2$O transformation in the conditions of the
inner envelope of IRC+10216. We consider an scenario given by
model MH (high density in order to have a short $\tau_{chem}$) and
the chemical reaction given by equation 9. This reaction can be
decomposed in several elementary reactions, for example
\begin{displaymath}
\begin{array}{rclc}
CO + H_2 & \stackrel{k_S}{\longrightarrow} & CH + OH & \quad (S) \\
OH + H_2 & \longrightarrow & H_2O + H & \\
CH + H_2 & \longrightarrow & CH_2 + H & \\
CH_2 + H_2 & \longrightarrow & CH_3 + H & \\
CH_3 + H_2 & \longrightarrow & CH_4 + H & \\
2 \times \quad H + H + H_2 & \longrightarrow & H_2 + H_2 & \\
\mbox{\boldmath $CO$ + $3$ $H_2$} & \stackrel{\mbox{\boldmath $k_G$}}{\mbox{\boldmath $\longrightarrow$}} & \mbox{\boldmath $CH_4$ $+$ $H_2O$} & \quad \mbox{\boldmath $(G)$} \\
\end{array}
\end{displaymath}
In any mechanism composed of elementary reactions, the rate of the
global reaction ($G$) will be given by the rate of the slowest
elementary reaction ($S$). In this case the slowest elementary
step will be the break of the CO bond, since it has the largest
activation barrier. The number of H$_2$O molecules produced or the
number of CO molecules destroyed in the global reaction, per unit
time and unit volume, is equal to the number of CO molecules
destroyed in the slowest elementary reaction, per unit time and
unit volume. We assign a rate constant $k_G$ for the global
reaction $G$, calculate the reverse rate constant via detailed
balance, and run the model for a CO,H$_2$,CH$_4$,H$_2$O system in
which only the global reaction $G$ and its reverse operate. Then
we vary $k_G$ until some significant fraction of CO is transformed
into H$_2$O and CH$_4$. Models for $k_G$ equal to 10$^{-42}$ and
10$^{-46}$ cm$^9$ s$^{-1}$ are shown in
Fig.~\ref{fig-quim-inn-vcte}d. It is seen that above a threshold
value of $\sim$10$^{-44}$ cm$^9$ s$^{-1}$ for $k_G$, the
CO$\rightarrow$H$_2$O transformation is significant. The rate of
destruction of CO molecules given by reaction $G$ is
$k_G$$\times$$x_{CO}$$\times$$n^4$ ($x_{CO}$ is the CO abundance
relative to H$_2$ and $n$ is the density of H$_2$) while the rate
of destruction of CO molecules by reaction $S$ is
$k_S$$\times$$x_{CO}$$\times$$n^2$. Since both rates have to be
equal, we arrive to $k_S$=$k_G$$\times$$n^2$. Taking 10$^{-44}$
for $k_G$ and a density of 10$^{10}$ cm$^{-3}$ (roughly the value
in the region 10-20 R$_*$ where the discussed process should take
place) results in $k_S$=10$^{-24}$ cm$^3$ s$^{-1}$. Therefore, we
need a bimolecular reaction similar to reaction $S$ with a rate
constant greater than 10$^{-24}$ cm$^3$ $s^{-1}$ for temperatures
lower than $\sim$700 K. There is no data in the literature about a
rate constant for a reaction such as $S$. Nevertheless, we can
estimate an upper limit for $k_S$ if we assume a maximum
temperature independent value of 10$^{-9}$ cm$^3$ s$^{-1}$ for the
reverse reaction and apply detailed balance. This results in
activation energies of $\sim$80,000 K for $k_S$, which for
temperatures below 700 K imply rate constants well below
10$^{-24}$ cm$^3$ s$^{-1}$. So we don't expect a chemical route
that is able to convert CO into H$_2$O in the inner envelope
within the gas phase and during an adiabatic expansion.\\

(ii) Such a CO$\rightarrow$H$_2$O process seems too slow, but
there is another O--bearing molecule which locks some important
amount of oxygen, i.e. SiO, which is expected to condense onto SiC
grains. How could H$_2$O abundance be affected by \emph{SiO
condensation} during the expansion?. When we consider such a
process in our model, H$_2$O abundance remains low. The reason is
that, as explained above, H$_2$O is related to SiO in such a way
that H$_2$O becomes abundant when SiO does, but not when SiO
depletes. Then, we consider in model MH (where SiO is formed with
a significant abundance) a speculative process that can be
summarized as follows: a SiO molecule condenses onto a SiC grain
and the Si atom incorporates into the grain lattice while the O
atom is released to the gas phase
\begin{displaymath}
SiO_{gas} + grain \rightarrow grain \cdots Si + O_{gas}
\end{displaymath}
We assume that 90\% of the SiO formed in the gas phase is
deposited onto SiC grains and that all that oxygen, formerly
contained in SiO, is released to the gas phase as atomic oxygen
(see Fig.~\ref{fig-quim-inn-vcte}e). It is seen that molecules
like H$_2$O and CO$_2$ increase their abundances. H$_2$O takes
efficiently the oxygen when it is in atomic form via reactions
R15+R11
\begin{displaymath}
\begin{array}{llcl}
R15 \qquad & O + H_2 & \rightarrow & OH + H \\
R11 \qquad & OH + H_2 & \rightarrow & H_2O + H \\
\end{array}
\end{displaymath}
although some important fraction of the atomic oxygen returns to
SiO from OH via reaction R10
\begin{displaymath}
\begin{array}{llcl}
R10 \qquad & OH + Si & \rightarrow & SiO + H \\
\end{array}
\end{displaymath}
and through CO$_2$ via reactions R14+R16
\begin{displaymath}
\begin{array}{llcl}
R14 \qquad & OH + CO & \rightarrow & CO_2 + H \\
R16 \qquad & CO_2 + Si & \rightarrow & SiO + CO \\
\end{array}
\end{displaymath}
It is known in materials science that oxidation of solid SiC
through a flow of O$_2$ at temperatures of $\sim$1000 K produces
solid SiO$_2$ and also CO and CO$_2$ molecules that diffuse trough
the lattice and are released to the gas phase (see \citealt{wan01}
and references therein). Therefore, some fraction of the oxygen
entering solid SiC grains begin to build the SiO$_2$ lattice while
the rest is released as CO and CO$_2$. In the region of the CSE
where SiC grain formation occurs, the temperatures are similar to
those described above but the conditions are somewhat different
since it is an oxygen-deficient environment. Nevertheless, if SiO
depletes on SiC grains, some fraction of the oxygen can be
processed in the grain and released as CO and CO$_2$. We have run
our model assuming that the oxygen of the SiO depleted on SiC
grains is released to the gas phase as CO$_2$ instead of atomic
oxygen. Both H$_2$O and SiO compete for the oxygen contained in
the CO$_2$ molecules through reactions R13 and R16 respectively
\begin{displaymath}
\begin{array}{llcl}
R13 \qquad & CO_2 + H_2 & \rightarrow & H_2O + CO \\
R16 \qquad & CO_2 + Si & \rightarrow & SiO + CO \\
\end{array}
\end{displaymath}
but the high activation energy of R13
prevents water from reaching a high abundance.\\

\begin{figure}
\includegraphics[angle=0,scale=.47]{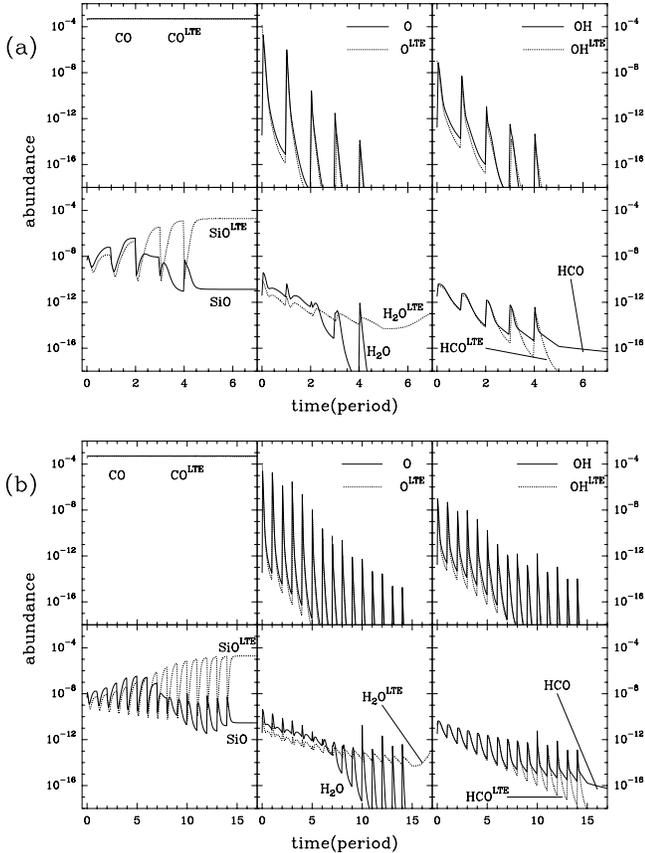}
\caption{Evolution of the abundances, relative to total number of
hydrogen nuclei, of some oxygen bearing molecules for a (a) 5
shocks history ($N_{shocks}$=5) and (b) 15 shocks history
($N_{shocks}$=15). The history starts at $r_0$ ($t$=0) and reaches
$r_c$ at time $N_{shocks}$$\times$$P$. The LTE abundances for the
physical conditions of the gas at each instant are shown as dotted
lines.} \label{fig-quim-inn-shoc}
\end{figure}

(iii) Now we will consider the effect that \emph{shocks} have on
oxygen chemistry. In Fig.~\ref{fig-quim-inn-shoc}
\notetoeditor{Fig.~\ref{fig-quim-inn-shoc} should appear with a 1
column width} we plot the evolution of some abundances as the gas
moves along the temperature and density profiles given by model MH
(see Fig.~\ref{fig-fis-inn-shoc}), in order to compare with
\citet{wil98} who used high density values. We also plot the LTE
abundances expected for the temperature and density profiles used,
so we can see how the abundances predicted by chemical kinetics
deviate from the LTE ones. Within the time interval associated to
one shock plus relaxation, the immediately post-shocked gas has
dissociated a certain fraction of CO releasing atomic oxygen which
has a large abundance during the first stages. This atomic oxygen
is progressively returned to CO during the relaxation time, but
also to OH which has an appreciable abundance during these first
stages. Reaction R11 (OH+H$_2$$\rightarrow$H$_2$O+H), which forms
water from OH, operates, but at these high temperatures the
reverse reaction immediately destroys the H$_2$O formed, returning
oxygen into OH and some time later into CO. Indeed, shocks are a
very good mechanism for dissociating CO and releasing atomic
oxygen. This CO$\rightarrow$O process occurs preferably during the
first shocks due to the high temperatures, which also make the
kinetics of the O$\rightarrow$H$_2$O process fast but, since water
is not thermodynamically abundant at high temperatures in a C-rich
gas (it is at low temperatures), other reactions are faster and
return the atomic oxygen to CO.

SiO is very affected by the high non-equilibrium chemistry driven
by shocks and ends this phase with a very low abundance, mainly
due to reaction R8, which was not included by \citet{wil98}. Note
that either considering 5 or 15 shocks to reach $r_c$ from $r_0$,
the abundances do not appreciably change. It is shown that this
mechanism decreases rather than enhances oxygen chemistry.

Concerning the H/H$_2$ ratio, shocks favor a large value since
H$_2$ is dissociated in the post-shocked gas (due to the high
temperatures reached) and depending on the density it can
regenerate during the relaxation time via the three body reactions
R1, R2 and R3 (with high densities, see
Fig.~\ref{fig-fis-inn-shoc}d) or freeze its abundance at some
moment during the relaxation (with low densities, see
Figs.~\ref{fig-fis-inn-shoc}e and \ref{fig-fis-inn-shoc}f). Again,
with a shocks mechanism, a high
density scenario is required to have hydrogen in molecular form.\\

In summary, O--bearing species are not produced efficiently by gas
phase chemistry in the inner envelope, neither considering an
expansion at constant velocity nor assuming shocks. Nevertheless,
the high water abundance obtained under the LTE assumption
suggests that grains may be playing an important role in the
recycling of oxygen from CO to H$_2$O.

\subsection{Oxygen chemistry in the outer envelope}

The chemistry in the outer envelope is mainly driven by the
photodissociation of molecules by the interstellar UV field
followed by rapid neutral-neutral reactions. This mechanism forms
species such as polyynes and cyanopolyynes from photodissociation
of parent species such as C$_2$H$_2$ and HCN \citep{mil94}.
Concerning oxygen, the photodissociation of CO occurs at the very
outer edge of the envelope due to self-shielding (much further
away than the photodissociation of C$_2$H$_2$ and HCN) so atomic
oxygen is abundant only at large radii in the CSE, where the
density has decreased considerably and reactions are very slow.
This fact makes oxygen chemistry to be much poor than carbon
chemistry, since oxygen keeps locked into CO along most of the
envelope. The inclusion of the isotopologue $^{13}$CO, $\sim$45
times less abundant than parent CO \citep{cer00}, which does not
self-shield due to its lower abundance, can enhance O abundance at
shorter radii. This allows for the formation of some O--bearing
species with moderate abundances through rapid neutral--neutral
reactions and radiative associations.

Ionic chemistry is triggered by cosmic rays ionization. Cosmic
rays can penetrate deeper than UV photons in the CSE since they
are not affected by opacity. Ions can react very rapidly via
ion-molecule reactions but the chemistry they can induce is
limited by the low abundance of ions in the CSE except at the very
outer edge (where for example all the carbon is converted into
C$^+$). In our model, ionic chemistry does not noticeably affect
the abundances of the O--bearing species formed except for SO,
whose abundance is greatly enhanced, and for HCO$^+$, produced
with a low abundance but enough for being detected due its large
dipole moment (4.07 D). Another important effect of ionic
chemistry is the fractionation of CO through the exchange reaction
R26
\begin{displaymath}
\begin{array}{llcl}
R26 \qquad & ^{13}C^+ + ^{12}CO & \rightarrow & ^{12}C^+ + ^{13}CO \\
\end{array}
\end{displaymath}
which is known to be exothermic by 35 K. Below this temperature
the reverse reaction does not operate and $^{12}$CO is effectively
transformed into $^{13}$CO. The net effect is a delay in the
destruction of $^{13}$CO: its abundance remains nearly constant
due to a steady state in which it is formed by the mentioned
exchange reaction while it is destroyed by UV photodissociation.

\begin{figure*}
\includegraphics[angle=-90,scale=.70]{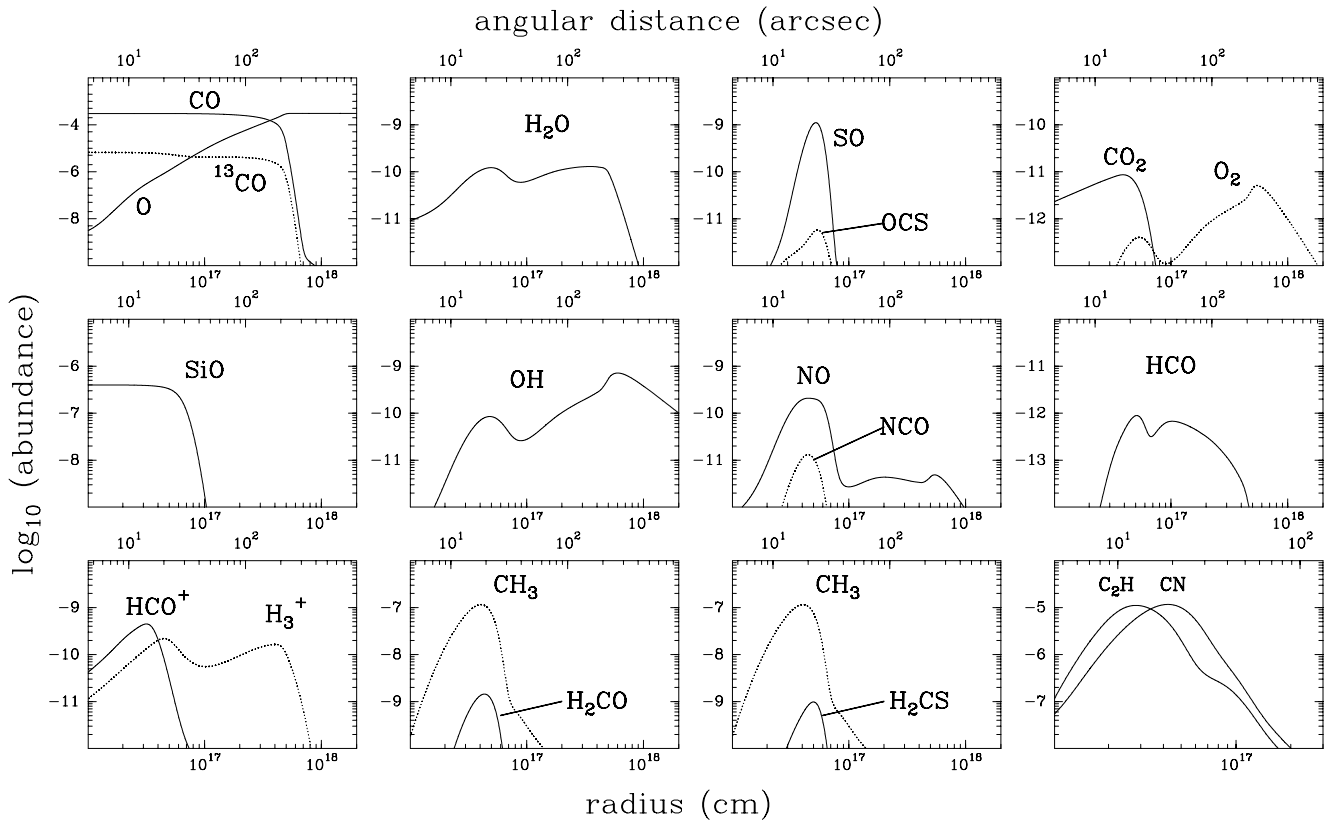}
\caption{Abundances, relative to total number of hydrogen nuclei,
for some oxygen bearing molecules in the outer expanding envelope
as given by chemical kinetics. C$_2$H and CN abundances are shown
in the bottom-right panel peaking at $\sim$15'' and $\sim$20''
respectively (see appendix B). Angular distance is given in the
top axis for an assumed distance to IRC+10216 of 150 pc.}
\label{fig-quim-out}
\end{figure*}

The abundances predicted in the outer envelope for CO, $^{13}$CO,
O, SiO and HCO$^+$ (usual O--bearing species in a C-rich star) are
shown in the left part of Fig.~\ref{fig-quim-out}
\notetoeditor{Fig.~\ref{fig-quim-out} should appear with a 2
column width}. The second column of panels show the abundances
predicted for H$_2$O, OH and H$_2$CO (the three "unexpected"
oxygen species), while the rest of panels to the right show the
abundances for some other O--bearing species. H$_2$CS abundance
distribution is also shown together with those of C$_2$H and CN,
peaking at $\sim$15'' and $\sim$20'' respectively (see appendix
B). We now describe individually the reactions that lead to their
abundances, which are expressed relative to total number of
hydrogen nuclei rather than relative to H$_2$:

-- \textbf{SiO} is formed in the inner envelope and its abundance
remains nearly constant throughout the outer envelope until it is
dissociated by UV photons (R41). Ionic chemistry destroys SiO at a
shorter radius by reacting with C$^+$ via R27
\begin{displaymath}
\begin{array}{llcl}
R27 \qquad & SiO + C^+ & \rightarrow & Si^+ + CO \\
\end{array}
\end{displaymath}
However, we point out that SiO could enter the outer envelope with
a much lower abundance due to condensation on grain surfaces.

-- \textbf{HCO$^+$} is produced by the same sequence of reactions
that forms it in dark clouds.
\begin{displaymath}
\begin{array}{llcl}
R44 \qquad & H_2 + Cosmic \quad Ray & \rightarrow & H_2^+ + e^- \\
R28 \qquad & H_2^+ + H_2 & \rightarrow & H_3^+ + H \\
R29 \qquad & H_3^+ + CO & \rightarrow & HCO^+ + H_2 \\
\end{array}
\end{displaymath}
First, H$_2$ is ionized by cosmic rays (R44), then H$_2^+$ reacts
with H$_2$ producing H$_3^+$ (R28), which in turn reacts rapidly
with CO through a proton transfer reaction (R29). The origin of
HCO$^+$ is different with respect to the rest of oxygen bearing
species commented in this section because its formation does not
require the photodissociation of CO but the presence of H$_3^+$,
the formation of which is related to cosmic rays. HCO$^+$ is
abundant at shorter radii than the rest of O--bearing species with
a peak abundance of 4.5$\times$10$^{-10}$ at 3$\times$10$^{16}$
cm, and a corresponding column density of 1.1$\times$10$^{12}$
cm$^{-2}$. The abundance profile sharply falls after its peak
mainly due to dissociative recombination R38 (electron density
becomes important beyond 5$\times$10$^{16}$ cm) and to a less
extent, due to proton transfer with HCN and C$_2$H$_2$ (reactions
R30 and R31 respectively).
\begin{displaymath}
\begin{array}{llcl}
R38 \qquad & HCO^+ + e^- & \rightarrow & CO + H \\
R30 \qquad & HCO^+ + HCN & \rightarrow & CO + HCNH^+ \\
R31 \qquad & HCO^+ + C_2H_2 & \rightarrow & CO + C_2H_3^+ \\
\end{array}
\end{displaymath}

-- \textbf{H$_2$O} formation in C-rich environments, where atomic
oxygen is available from the photodissociation of CO, is possible
through the route of reaction R15 followed by reaction R11.
\begin{displaymath}
\begin{array}{llcl}
R15 \qquad & O + H_2 & \rightarrow & OH + H \\
R11 \qquad & OH + H_2 & \rightarrow & H_2O + H \\
\end{array}
\end{displaymath}
This is the mechanism that forms water in the C-rich
protoplanetary nebula CRL618 (see \citealt{cer04}), where
temperatures in the inner slowly expanding envelope are $\sim$300
K. However, in the outer envelope of IRC+10216 such high
temperatures are not expected and this route does not work, since
both reactions R15 and R11 have activation energies that result in
negligible rates. In our model, water forms through the radiative
association between atomic oxygen and molecular hydrogen (R45).
\begin{displaymath}
\begin{array}{llcl}
R45 \qquad & O + H_2 & \rightarrow & H_2O + h\nu \\
\end{array}
\end{displaymath}
To our knowledge, neither experimental nor theoretical studies
have made an estimation of this rate. We have assumed a
temperature independent rate constant of 10$^{-18}$ cm$^3$
s$^{-1}$, typical of radiative associations between neutral
species. Reaction R45 implies breaking the H$_2$ bond so that
atomic oxygen can insert in, which will result on some activation
energy. However, similar reactions, e.g. H$_2$+C \citep{let00} and
H$_2$+CH \citep{bro97} take values 10$^{-17}$-10$^{-18}$ cm$^3$
s$^{-1}$. Water is produced over an extended region from
4$\times$10$^{16}$ to 6$\times$10$^{17}$ cm with a peak abundance
of 10$^{-10}$ and a column density of 4$\times$10$^{11}$
cm$^{-2}$. These values are far below the abundance estimated from
the observations of SWAS (2-12$\times$10$^{-7}$, \citealt{mel01})
or ODIN (1.2$\times$10$^{-6}$, \citealt{has06}) by at least three
orders of magnitude. We will discuss in \S 4.3 if water originated
in this way can or cannot explain the observations.

-- \textbf{OH} is produced from the photodissociation of water
(R42) and from atomic oxygen through both the bimolecular reaction
R17
\begin{displaymath}
\begin{array}{llcl}
R17 \qquad & O + NH_2 & \rightarrow & OH + NH \\
\end{array}
\end{displaymath}
and the radiative association between O and H (R36). As the gas
expands, reaction R36 becomes slower due to the decrease in
density although it is compensated by the increase in atomic
oxygen abundance. The result is an increase of OH abundance only
at the very outer envelope, shown in Fig.~\ref{fig-quim-out} as a
peak at 6$\times$10$^{17}$ cm. In fact, the O--bearing molecules
formed by radiative associations show an extended distribution
(reactants such as atoms increase their abundances while
expanding) whereas those formed by bimolecular reactions show a
peak distribution (reactions only proceed in a shell where
reactants are abundant before they are photodissociated). OH could
also be produced in reactions of O with several abundant
hydrocarbons (e.g. C$_n$H$_2$, C$_n$H, HC$_{2n+1}$N) through H
abstraction. However, the reactions of atomic oxygen with
closed-shell species such as C$_2$H$_2$, C$_4$H$_2$ and HCN have
activation barriers whereas in reactions with radicals the H
abstraction channel seems to proceed slowly. For example, the
reaction O+CH$\rightarrow$C+OH has been studied by \citet{mur86}
who estimated a negligible reaction rate at low temperatures while
the reaction of O with C$_{2}$H produces mainly CO and CH. OH is
predicted with a very extended distribution from
4$\times$10$^{16}$ to 10$^{18}$ cm with a peak abundance of
7$\times$10$^{-10}$ at the very outer edge and a column density of
2$\times$10$^{11}$ cm$^{-2}$. \citet{for03} have derived an OH
abundance of 4$\times$10$^{-8}$ assuming that all OH is produced
from the photodissociation of H$_2$O, which is formed through
comet evaporation in the inner envelope. The spatial distribution
they obtain ranges from 3$\times$10$^{16}$ cm to
7$\times$10$^{16}$ cm. If OH follows the more extended radial
distribution calculated by us, then the beam filling factors and
physical conditions of the emitting gas are very different from
those assumed by \citet{for03} and we could expect a considerable
decrease of the OH estimated abundance.

-- \textbf{H$_2$CO} is produced through reaction R18
\begin{displaymath}
\begin{array}{llcl}
R18 \qquad & O + CH_3 & \rightarrow & H_2CO + H \\
\end{array}
\end{displaymath}
the source of CH$_3$ being the photodissociation of CH$_4$ (R43)
while atomic oxygen is provided by CO and $^{13}$CO
photodissociation. The rate for R18 is well known for a
temperature range of 259-2500 K in which it does not show a
temperature dependence, suggesting a similar value at low
temperatures. The peak abundance is 1.4$\times$10$^{-9}$, reached
at 4$\times$10$^{16}$ cm. The abundance derived by \citet{for04}
from millimeter observations of pure rotational transitions is a
factor 5 higher than the one we get with the model. We discuss in
\S 4.3 the compatibility of the abundance predicted by our model
with the lines observed.

-- \textbf{SO} can be formed from OH+S (R19) and from O+SH (R20).
\begin{displaymath}
\begin{array}{llcl}
R19 \qquad & OH + S & \rightarrow & SO + H \\
R20 \qquad & O + SH & \rightarrow & SO + H \\
\end{array}
\end{displaymath}
For R19 we take the rate constant measured at 298 K. The rate of
reaction R20 has been measured at 298 K and at high temperatures
(1100-2000 K) and does not show temperature dependence, so that we
assume the same value at low temperatures. In our model the
abundance of OH is low in most of the CSE. On the other hand, SH
abundance is relatively high due to a sequence of ion-molecule
reactions that begin with the charge transfer of C$^+$ to SiS
(R32). The SiS$^+$ formed this way reacts with H via reaction R33
and produces SH.
\begin{displaymath}
\begin{array}{llcl}
R32 \qquad & SiS + C^+ & \rightarrow & SiS^+ + C \\
R33 \qquad & SiS^+ + H & \rightarrow & SH + Si^+ \\
\end{array}
\end{displaymath}
The peak abundance predicted for SO is 1.1$\times$10$^{-9}$ with a
column density of 4.7$\times$10$^{11}$ cm$^{-2}$. However, we
point out that SO abundance is greatly influenced by H abundance,
for which we have taken an upper limit given by \citet{bow87}. A
process of H$_2$ formation on grains surfaces, not considered in
this model, could reduce the abundance of atomic H, thus
decreasing the amount of SO in the outer envelope.

-- \textbf{OCS} forms by radiative association of S and CO (R37).
The predicted peak abundance is 6$\times$10$^{-12}$ with a quite
low column density of 7.4$\times$10$^{9}$ cm$^{-2}$.

-- \textbf{NO} is formed by the bimolecular reaction R21 involving
atomic oxygen and NH, which in turn comes from photodissociation
of NH$_3$.
\begin{displaymath}
\begin{array}{llcl}
R21 \qquad & O + NH & \rightarrow & NO + H \\
\end{array}
\end{displaymath}
The rate for this reaction is a temperature independent estimation
for the 250-3000 K range and is assumed to apply also at low
temperatures. The predicted abundance is 2$\times$10$^{-10}$ with
a column density of 2$\times$10$^{11}$ cm$^{-2}$, although its low
dipole moment (0.159 D) makes difficult its detection.

-- \textbf{NCO} forms by reaction R22 the rate of which is based
on an experimental measurement at low pressure and 292 K and the
assumption that formation of NCO is the main channel.
\begin{displaymath}
\begin{array}{llcl}
R22 \qquad & CN + OH & \rightarrow & NCO + H \\
\end{array}
\end{displaymath}
Its predicted abundance is 10$^{-11}$. The relatively low dipole
moment (0.6 D), the large partition function and the low predicted
abundance will make very difficult its detection.

-- \textbf{H$_2$CS} abundance is also shown in
Fig.~\ref{fig-quim-out}. Although it is not an O--bearing species,
it is formed through the sulfur analogous reaction that forms
H$_2$CO (R23).
\begin{displaymath}
\begin{array}{llcl}
R23 \qquad & S + CH_3 & \rightarrow & H_2CS + H \\
\end{array}
\end{displaymath}
The rate of this reaction has not been measured and therefore it
is assumed equal to the oxygen analogous case. Atomic sulfur is
necessary, instead of oxygen, and it is provided by the
photodissociation of SiS and CS. Ionic chemistry slightly enhances
H$_2$CS abundance since atomic S is released at short radii
through ion-molecule reactions such as R34
\begin{displaymath}
\begin{array}{llcl}
R34 \qquad & SiS + S^+ & \rightarrow & SiS^+ + S \\
\end{array}
\end{displaymath}
and because more chemical routes for thioformaldehyde formation
appear, e.g. from H$_3$CS$^+$ through dissociative recombination
(R39).
\begin{displaymath}
\begin{array}{llcl}
R39 \qquad & H_3CS^+ + e^- & \rightarrow & H_2CS + H \\
\end{array}
\end{displaymath}
The peak abundance is 10$^{-9}$ with a distribution similar to
that of formaldehyde. The predicted abundance profile produces
line shapes in reasonable agreement with observations (see below).

-- \textbf{CO$_2$} is formed through reaction R35, which involves
SiO$^+$ that is formed with an abundance of $\sim$10$^{-11}$ from
several ion-molecule reactions.
\begin{displaymath}
\begin{array}{llcl}
R35 \qquad & SiO^+ + CO & \rightarrow & CO_2 + Si^+ \\
\end{array}
\end{displaymath}
We point out that CO$_2$ could be also a product of evaporation of
cometary ices and will have a large abundance in the cometary
scenario. However, an ISO spectrum of IRC+10216 in the
mid-infrared with a very large signal-to-noise ratio \citep{cer99}
only shows C$_2$H$_2$ and HCN lines in the 13-16 $\mu$m range and
no evidence for CO$_2$.

-- \textbf{O$_2$} formation is strongly related to the presence of
OH from which is formed through reaction R24.
\begin{displaymath}
\begin{array}{llcl}
R24 \qquad & O + OH & \rightarrow & O_2 + H \\
\end{array}
\end{displaymath}
Its spatial distribution follows that of OH with an abundance 2
orders of magnitude smaller.

-- \textbf{HCO} is produced with a quite low abundance
($\sim$10$^{-12}$) mainly from atomic oxygen by reaction R25.
\begin{displaymath}
\begin{array}{llcl}
R25 \qquad & O + CH_2 & \rightarrow & HCO + H \\
\end{array}
\end{displaymath}
Dissociative recombination of H$_2$CO$^+$ (R40) also contributes
to HCO formation.

\subsection{Comparison with observations}

In order to compare with available observations we have performed
Monte Carlo radiative transfer calculations for H$_2$CO, H$_2$O,
HCO$^+$, SO and H$_2$CS with the code described in \citet{gon93}.
The CSE has been simulated by a spherically distributed expanding
gas with the radial profiles for density, temperature and
abundance of each species taken from the chemical model. Line
profiles and intensities for several transitions are then
generated after convolution of the radiative transfer results with
the main beam of the selected telescope.

\begin{figure*}
\includegraphics[angle=-90,scale=.75]{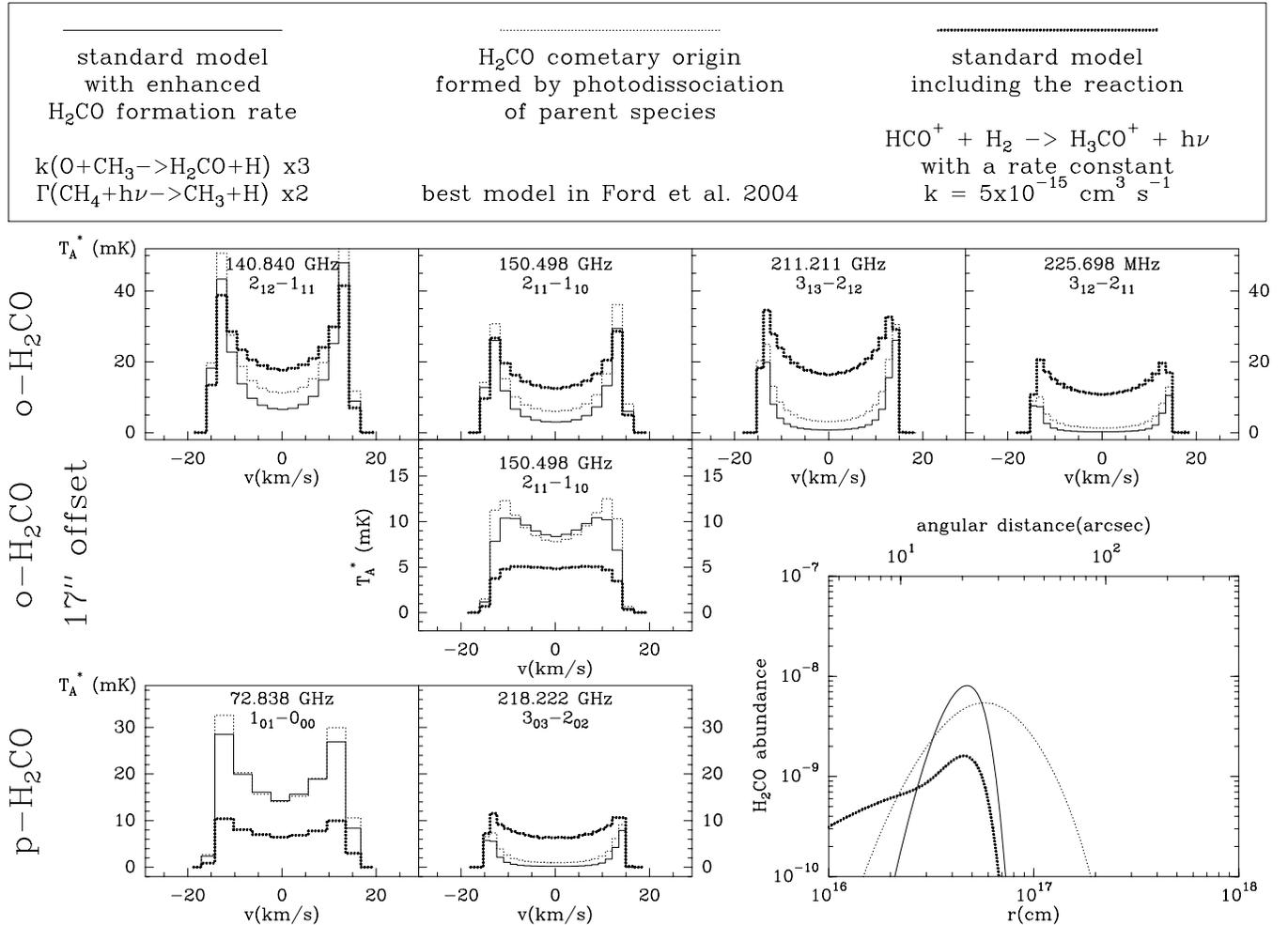}
\caption{Calculated line intensities and profiles for pure
rotational transitions of ortho-H$_2$CO and para-H$_2$CO. The
lines should be compared with the observations reported by
\citet{for04}. IRAM-30m beam size has been considered and the
lines have been smoothed to a resolution of 1 MHz. The H$_2$CO
abundances, relative to total number of H nuclei, are plotted in
the bottom-right panel. --Solid lines: H$_2$CO abundance resulting
from the chemical model with two parameters slightly changed: R18
rate constant increased in a factor 3 and the CH$_3$ branching
ratio in methane photodissociation increased from 15\% to 30\%.
--Faint dotted lines: H$_2$CO produced from photodissociation of
an unknown parent species, which would be released by sublimation
of the mantles of comets, assumed to be present in the inner
envelope. Best model in \citet{for04}. --Hard dotted lines:
H$_2$CO abundance resulting from the chemical model including
reaction R46 in order to increase formaldehyde abundance in the
inner regions.} \label{fig-spec-h2co}
\end{figure*}

-- \textbf{H$_2$CO}. \citet{for04} have observed four pure
rotational lines of ortho-H$_2$CO: $2_{1,2}-1_{1,1}$ (140.840
GHz), $2_{1,1}-1_{1,0}$ (150.498 GHz), $3_{1,3}-2_{1,2}$ (211.211
GHz) and $3_{1,2}-2_{1,1}$ (225.698 GHz) with antenna temperatures
of $\sim$30 mK, $\sim$20 mK, $\sim$20 mK and $\sim$20 mK
respectively; one line of para-H$_2$CO: $3_{0,3}-2_{0,2}$ (218.222
GHz) with $T_{A}^{*}$$\sim$20 mK; and failed to detect the low
excitation transition of para-H$_2$CO $1_{0,1}-0_{0,0}$ (72.838
GHz) at a sensitivity of $\sim$10 mK. They also conducted 17''
offset observations for the $2_{1,1}-1_{1,0}$ line with an average
line intensity reduced by a factor 3-4, which nevertheless
indicates an extended H$_2$CO distribution. We have used the
abundance profile obtained from our chemical model to predict the
expected line intensities in order to see if the formaldehyde
formed through these chemical routes can account for the observed
lines. The collisional coefficients, needed for solving the
statistical equilibrium, are taken from \citet{gre91} (corrected
for H$_2$ as collider instead of He) and the normal ortho-to-para
ratio of 3:1 has been assumed.

The resulting lines are less intense than the observed ones by a
factor 4-10 depending on the line. Nevertheless, one has to keep
in mind that the abundances resulting from such a simple chemical
model are affected by uncertainties due to missing reactions,
uncertain rates and also errors in physical parameters (kinetic
temperature, mass loss rate, etc). Formaldehyde is formed from O
and CH$_3$ by reaction R18
\begin{displaymath}
\begin{array}{llcl}
R18 \qquad & O + CH_3 & \rightarrow & H_2CO + H \\
\end{array}
\end{displaymath}
so the abundances of the two reactants together with the rate of
R18 directly affect the abundance that H$_2$CO may reach. The rate
constant for reaction R18 is known down to 259 K, but not for the
much lower temperatures that are present in the outer envelope of
IRC+10216. We could expect a higher rate constant at very low
temperatures, a behavior experimentally observed in many other
radical-molecule reactions without activation energy (see
\citealt{smi04} and references therein). Another important
parameter is the branching ratio for CH$_3$ production in methane
photodissociation by the interstellar UV field, for which
\citet{let00} give a 15\%. For the solar UV field values up to
44\% are commonly used (see the chemical model of Saturn's
atmosphere by \citealt{mos00}). Although the energy distribution
of the interstellar and solar UV field are somewhat different, we
could expect larger uncertainties in the former one. When we
increase R18 rate constant by a factor of 3 and the CH$_3$
branching ratio by a factor of 2, the resulting H$_2$CO abundance
(see solid line in right-down panel of Fig.~\ref{fig-spec-h2co}
\notetoeditor{Fig.~\ref{fig-spec-h2co} should appear with a 2
column width}) increases by a factor of $\sim$6 with respect to
the former model (that of Fig.~\ref{fig-quim-out}). A comparison
of the spectral lines resulting from such abundance profile with
the lines observed by \citet{for04} give us several conclusions:
(1) the line intensities agree quite well for the ortho-H$_2$CO
lines whereas for the para-H$_2$CO the agreement is worse: the
72.838 GHz line is predicted a factor $\sim$3 more intense than
the observed upper limit and the 218.222 GHz line is predicted a
factor 2-3 less intense than the observation. Uncertainties in the
collisional coefficients, ortho-to-para ratio or in the
temperature and density profiles used could explain the
discrepancies between model and observations. (2) The ratio of the
integrated intensity between the 17'' offset position to the
centered one in the 150.498 GHz line is 0.87, in contrast with the
value of 0.4 obtained from the observations. Sources for the
disagreement can be errors in the pointing, the uncertainty in the
distance to the star or the depart from spherical symmetry in
H$_2$CO emission. (3) One last aspect to point out is the
different shapes of the observed and predicted 1.3 mm lines of
ortho-H$_2$CO (at 211.211 and 225.698 GHz). The predicted lines
have pronounced double-peaked shapes with almost no emission at
zero velocity, as corresponds to a shell with a certain size
(diameter of $\sim$40'' in the model) observed with a smaller
telescope beam ($\sim$11'' for IRAM-30m at those frequencies). On
the other hand the observed lines show substantial emission at
zero velocity and a less pronounced double-peaked profile. Neither
an error of up to 4'' in the pointing observations nor a larger
distance to the star of up to 200 pc (more will make IRC+10216 to
be overluminous for its type according to \citealt{cro97}) will
appreciably change the predicted line profiles.

\citet{for04} interpret the presence of formaldehyde as a direct
product from the photodissociation of an unknown parent molecule
released in the inner envelope by cometary sublimation. Such
interpretation is supported by the observation of extended
emission of formaldehyde toward the comet P/Halley \citep{mei93}.
\citet{for04} constructed a simple model with the following
parameters: initial abundance of the parent species relative to
H$_2$ of 8$\times$10$^{-8}$, unattenuated photodissociation rate
of the parent species producing H$_2$CO of
$A_p$=1.6$\times$10$^{-10}$ s$^{-1}$ and unattenuated
photodissocation rate of formaldehyde
$A_{H_2CO}$=7.81$\times$10$^{-10}$ s$^{-1}$, which reproduced the
observed line intensity of the 150.498 GHz line and the ratio
($\int T_A^* dv_{off}$)/($\int T_A^* dv_{cent}$) of 0.4 for this
line. We have run a model with the same parameters to obtain the
abundance distribution and the expected line profiles (light
dotted lines in Fig.~\ref{fig-spec-h2co}). It is seen that the
abundance distribution is more extended than the previous one
although the 1.3 mm line profiles of ortho-H$_2$CO still show an
emission at zero velocity too low compared with the observations.
The abundance distribution of H$_2$CO with a cometary origin is
somewhat similar to the one we get with the chemical model,
because in both cases formaldehyde is produced in the region where
interstellar UV photons penetrate.

\begin{figure*}
\includegraphics[angle=-90,scale=.70]{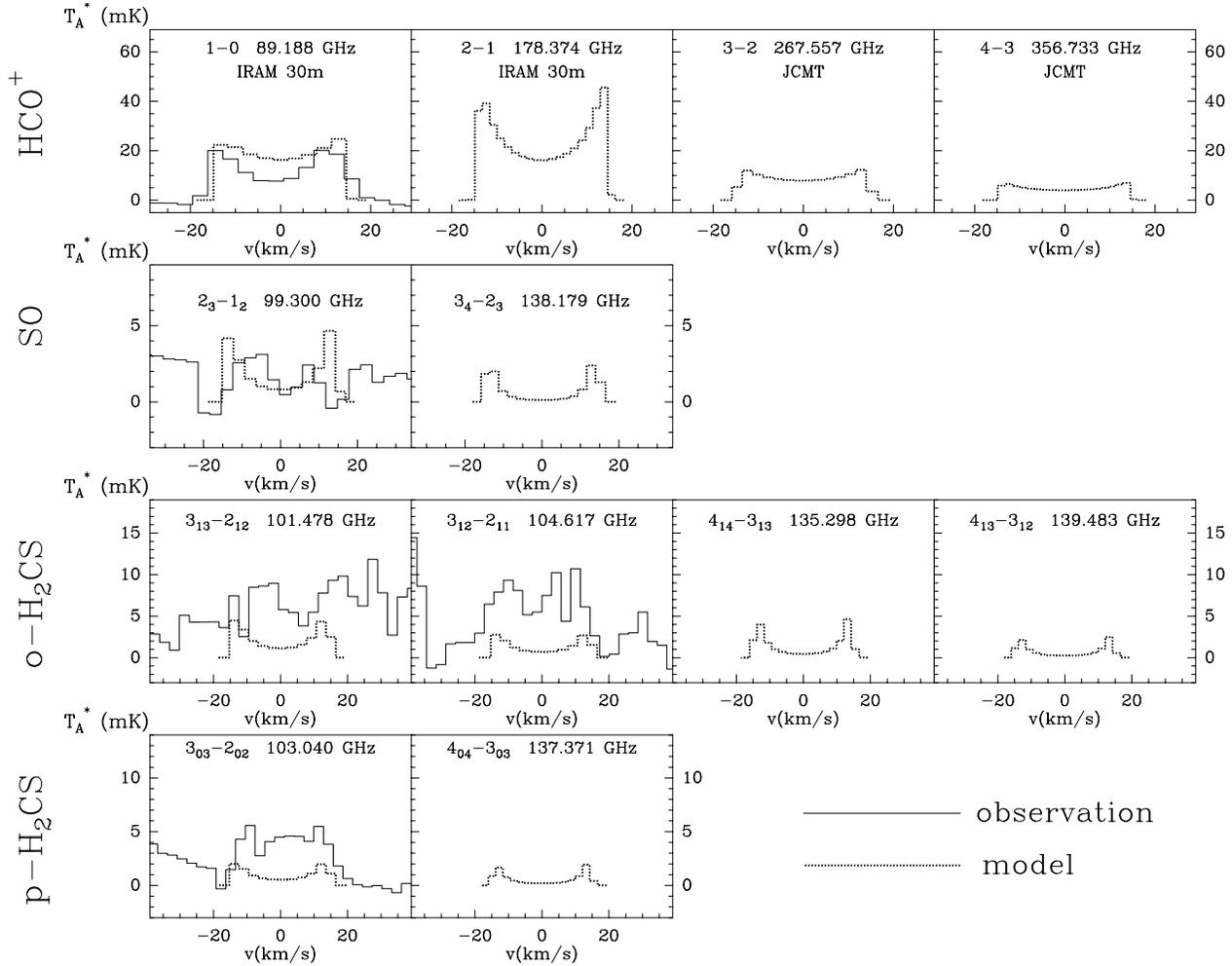}
\caption{Calculated line intensities and profiles for pure
rotational transitions of HCO$^+$, SO, ortho-H$_2$CS and
para-H$_2$CS. All lines have been smoothed to a resolution of 1
MHz except the 3-2 and 4-3 of HCO$^+$ which have been smoothed to
a 2 MHz resolution. All lines are calculated for IRAM 30m
telescope except the 3-2 and 4-3 which have been calculated for
the JCMT telescope. The abundance profiles used are given by the
chemical model (those of Fig.~\ref{fig-quim-out}) except that of
HCO$^+$, which has been decreased by a factor 2 for matching the
$J$=1-0 line observed. The model is plotted as dotted lines and
the $\lambda$ 3 mm observations from \citet{cer06} as solid
lines.} \label{fig-spec-other}
\end{figure*}

From the above considerations it follows that an increase of the
formaldehyde abundance in the inner regions of the CSE is needed
for producing some emission at zero velocity and matching the
observations. Formation of H$_2$CO in inner regions can be
explained by hydrogenation of CO on grain surfaces followed by
photodesorption (species such as CH$_4$, NH$_3$ and H$_2$S are
thought to be formed by similar processes). Since this possibility
is very complex to model, we have investigated if some missing gas
phase reactions can enhance formaldehyde abundance at short radii.
At this point it is worth to note that HCO$^+$ has an abundance
distribution peaking closer to the star than the one of H$_2$CO
(see Fig~\ref{fig-quim-out}) and that produces double-peaked
spectral lines but with substantial emission at zero velocity (see
Fig.~\ref{fig-spec-other} \notetoeditor{Fig.~\ref{fig-spec-other}
should appear with a 2 column width}). As discussed for HCO$^+$
formation, ionic chemistry is initiated by cosmic rays at shorter
distances than photochemistry does since the CSE is not opaque to
cosmic rays. Formaldehyde can also be formed from H$_2$CO$^+$
(through charge transfer with several species) and H$_3$CO$^+$
(through dissociative recombination) but in our model these routes
are not important since both H$_2$CO$^+$ and H$_3$CO$^+$ are not
abundant enough. Looking at exothermicity we can examine possible
missing reactions that can contribute to the formation of these
two species (heats of formation have been taken from NIST
Chemistry Webbook\footnote{The NIST Chemistry Webbook is available
on the World Wide Web at http://webbook.nist.gov/chemistry/.} and
\citealt{let00}). One possibility is the reaction
\emph{H$_3^+$+CO} for which the only exothermic channels are the
well known proton transfer giving HCO$^+$ and the radiative
association which produces H$_3$CO$^+$. Another possibility is the
reaction \emph{HCO$^+$+H$_2$} but all the investigated channels
are highly endothermic due to the high stability of both reactants
and only the radiative association producing H$_3$CO$^+$ is
exothermic by 123 kJ/mol. We find that including the reaction
HCO$^+$+H$_2$ $\rightarrow$ H$_3$CO$^+$+h$\nu$ with a rate
constant of 5$\times$10$^{-15}$ cm$^3$ s$^{-1}$ (R46), the
formaldehyde abundance is enhanced at the inner regions producing
rotational lines in agreement with the observations, i.e. with
significant emission at zero velocity (strong dotted lines in
Fig.~\ref{fig-spec-h2co}). All the lines except the one at 218.222
GHz of para-H$_2$CO (predicted a factor 2 less intense than
observed) agree very well with the observations in intensity and
profile. The required rate constant is typical of other
ion-neutral radiative associations \citep{let00}.

Finally, is it worth to note that three of the four unidentified
lines in \citet{for04} can be assigned as follows: U150 is the
7$_{1,7}$-6$_{1,6}$ transition of Si$^{13}$CC at 150385.281 MHz
\citep{cer91,cer00}; U218b is the $N$=23-22 $J$=45/2-43/2
transition of CC$^{13}$CCH at 218103.281 MHz \citep{cer00}; and
U218a is one component of the $^2\Pi_{1/2}$ $J$=45/2-43/2
transition in the $\nu_7$=1 vibrational state of C$_4$H at
218287.460 MHz
\citep{gue87b,yam87,cer00}.\\

-- \textbf{H$_2$O}. The evidence for water presence in IRC+10216
comes from the detection of the ortho transition
1$_{1,0}$-1$_{0,1}$ at 556.936 GHz with the telescopes SWAS
($T_A^*$$\sim$20 mK, \citealt{mel01}) and ODIN ($T_{MB}$$\sim$50
mK, \citealt{has06}). In order to model the radiative transfer of
water throughout the CSE we consider the lowest lying 8 rotational
levels in the vibrational ground state level
($\nu_1$,$\nu_2$,$\nu_3$)=(0,0,0), in the first excited bending
mode (0,1,0) and in the first excited asymmetric stretching mode
(0,0,1). The ortho-H$_2$O--He collisional rates from
\citet{gre93}, corrected for H$_2$ as collider, are adopted for
all the transitions within the same vibrational level, whereas for
ro-vibrational transitions collisions are not considered
important.

Excitation to the $\nu_2$=1 state occurs by absorption of
$\lambda$$\sim$6 $\mu$m photons whereas excitation to the
$\nu_3$=1 state corresponds to $\lambda$$\sim$3 $\mu$m. Photons at
these two wavelengths are abundant within the CSE due to the
presence of the central star and dust, so these excited
vibrational levels can be easily populated. Furthermore, since our
chemical model predicts water to be abundant in the outer CSE,
where densities range from 10$^5$ to 10$^2$ cm$^{-3}$
(n$_{crit}$$\sim$10$^9$ cm$^{-3}$ for the 556.936 GHz transition),
collisional excitation is not effective to populate the 1$_{1,0}$
state and the dominant excitation mechanism is radiative pumping
to the $\nu_2$=1 and $\nu_3$=1 vibrational excited states followed
by radiative decay to several rotational levels of the vibrational
ground state, included the 1$_{1,0}$ state.

\begin{figure*}
\includegraphics[angle=-90,scale=.73]{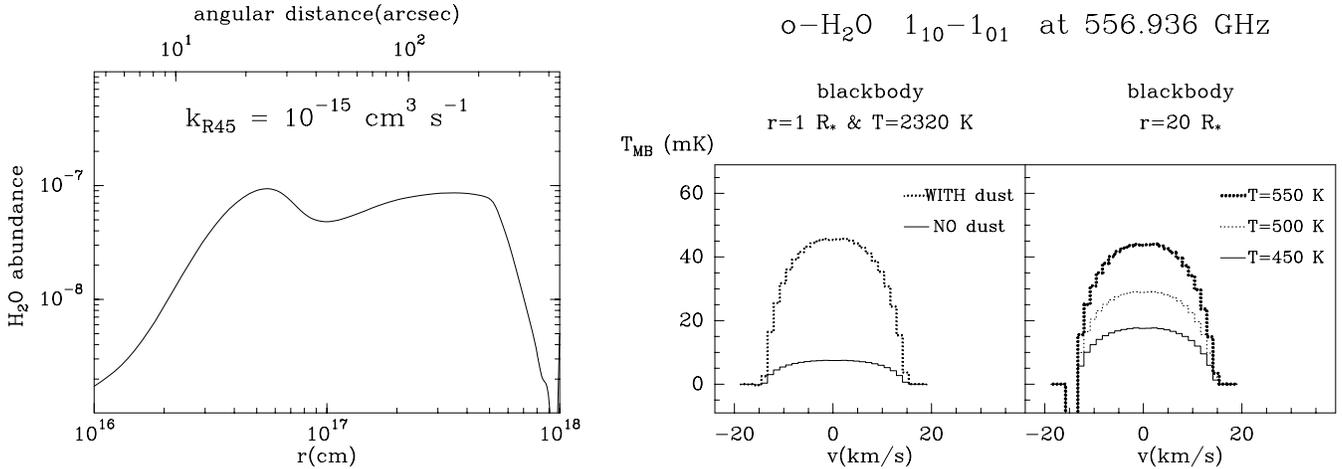}
\caption{--Left: Water vapor abundance, relative to total number
of H nuclei, when assuming a rate constant for R45 of 10$^{-15}$
cm$^3$ s$^{-1}$. --Right: calculated line intensities and profiles
for the 1$_{1,0}$-1$_{0,1}$ transition of ortho-H$_2$O for
different IR fluxes. ODIN beam size ($\sim$2') has been considered
and the lines have been smoothed to a resolution of 1.3 km/s.}
\label{fig-spec-water}
\end{figure*}

In our chemical model water is formed through the radiative
association R45
\begin{displaymath}
\begin{array}{llcl}
R45 \qquad & O + H_2 & \rightarrow & H_2O + h\nu \\
\end{array}
\end{displaymath}
with an assumed rate constant of 10$^{-18}$ cm$^3$ s$^{-1}$. As
noted in \S 4.2, the abundance reached with such a rate constant
is not enough to explain the observed 556.936 GHz line. Since the
rate constant for this reaction is not known and water abundance
approximately scales with this rate, we have fitted R45 rate to
get a 556.936 GHz line intensity in agreement with the
observations. We find that it is necessary to increase the rate
constant up to $\sim$10$^{-15}$ cm$^3$ s$^{-1}$. The peak
abundance obtained then is $\sim$10$^{-7}$ as shown in
Fig.~\ref{fig-spec-water} \notetoeditor{Fig.~\ref{fig-spec-water}
should appear with a 2 column width}. With such an abundance
profile, we have performed Monte Carlo radiative transfer
calculations. We show in Fig.~\ref{fig-spec-water} the lines
predicted for ODIN in different scenarios in which the fluxes at
$\lambda$=3 and 6 $\mu$m vary. On the left side the central star
has been considered as a blackbody at $T$=$T_*$ with and without
dust throughout the CSE. We consider carbonaceous grains with
sizes between 0.01 and 1 $\mu$m and a gas-to-dust mass ratio of
500 \citep{kna85}. It is seen that the presence of dust enhances
the 557 GHZ line intensity since grains absorb the photons from
the star (its maximum is at $\lambda$=2 $\mu$m) and reemits at
longer wavelengths, thus enhancing the flux at $\lambda$=6$\mu$m
and exciting the bending mode of H$_2$O. In fact, the IR spectrum
of IRC+10216 as observed by ISO \citep{cer99} peaks at 10 $\mu$m
due to the presence of dust and approximately corresponds to a
blackbody of radius 20 R$_*$ with a temperature of $\sim$500 K. We
can therefore simulate the IR flux that water in the outer
envelope would receive by putting such a blackbody instead of the
star. In Fig.~\ref{fig-spec-water} we see how the 556.936 GHz line
intensity increases when the flux at $\lambda$ 6 $\mu$m is
enhanced by considering a blackbody with increasing temperature.
These radiative transfer models show the importance of the IR
pumping of the excited vibrational levels of H$_2$O for the
intensities of pure rotational lines in the ground vibrational
state.

In summary, assuming a rate constant of 10$^{-15}$ or a few
10$^{-16}$ cm$^3$ s$^{-1}$ for R45 we can reproduce the water line
observed by ODIN. Now the question is how reasonable is such a
high rate constant for a radiative association between neutrals.
This reaction is spin-forbidden according to Wigner-Witmer rules
since the potential energy surfaces (PES) of reactants and
products in their ground electronic states,
H$_2$($^1$$\Sigma$)+O($^3$P)$\rightarrow$H$_2$O($^1$A)+h$\nu$, are
not connected adiabatically. For the reaction to proceed, an
intersystem crossing between PES with different total electronic
spin has to occur during the reaction. A spin-forbidden reaction
is likely to have a small rate constant, however some other
reactions in which reactants and products do not adiabatically
correlate have been found to be very rapid. For example the
bimolecular reaction:
C($^3$P)+C$_2$H$_2$($^1$$\Sigma$)$\rightarrow$C$_3$($^1$$\Sigma$)+H$_2$($^1$$\Sigma$)
may have a rate constant of several 10$^{-10}$ cm$^3$ s$^{-1}$ at
very low temperatures \citep{cla02} while the radiative
association
HS$^+$($^3$$\Sigma$)+H$_2$($^1$$\Sigma$)$\rightarrow$H$_3$S$^+$($^1$A)+h$\nu$
has a rate constant of 7$\times$10$^{-16}$ cm$^3$ s$^{-1}$ at 80 K
\citep{her89}.

A high rate constant for reaction R45 would also have consequences
for the chemistry in dark clouds. If k$_{R45}$=10$^{-15}$ cm$^3$
s$^{-1}$, and with a typical density of 10$^4$ cm$^{-3}$, the
chemical time scale to produce water from atomic oxygen would be
as short as 3$\times$10$^3$ years. Chemical models without this
radiative association predict H$_2$O abundances somewhat higher
than the observational upper limits imposed by SWAS \citep{rob02}.
Nevertheless, depletion on grains at late times would reduce the
water vapor abundance so that models could match the observations.
In diffuse clouds, with lower visual extinctions ($A_v$$\sim$1)
and densities ($n_H$$\sim$100 cm$^{-3}$), reaction R45 would also
produce water in time scales of 3$\times$10$^5$ years, with an
abundance relative to H$_2$ of $\sim$3$\times$10$^{-7}$ in steady
state\footnote{The H$_2$O abundance reaches the steady state when
the formation and destruction rates become equal. If we assume
that reactions R45 and R42 are the main formation and destruction
routes, then $n$(H$_2$O)/$n$(H$_2$) = [$k_{R45}$ $x_O$
$n_{H_2}$]/[$G_0$ 5.9$\times$10$^{-10}$ exp($-7.1$ $A_v$)], where
$x_O$ is the O abundance relative to H$_2$ and $G_0$ is the UV
field relative to the interstellar standard one (see appendix B).
If we take $A_v$=1, $G_0$=1, $x_O$=6$\times$10$^{-4}$,
$k_{R45}$=10$^{-15}$ cm$^3$ s$^{-1}$ and $n$(H$_2$)=$n_H$/2=100/2
cm$^{-3}$ (assuming hydrogen is mainly molecular), we get
$n$(H$_2$O)/$n$(H$_2$) = 3$\times$10$^{-7}$.}. This value is 1--2
orders of magnitude higher than the scarce observational
estimations available \citep{spa98,neu02}. Nevertheless, in
diffuse clouds a significant fraction of hydrogen is not in
molecular form, therefore the rate of water formation through
reaction R45 would be lower than our estimation. Furthermore,
since the observations of interstellar water ice bands do not
discriminate between different types of clouds, dense or diffuse,
present along the line of sight (e.g. \citealt{sch98,mon01}), it
is highly uncertain whether the amount of water as solid ice is
important or not.

There are other clues, apart from the 557 GHz line, that will help
in understanding the water origin in this late-type C-rich star:
(1) If water is formed in the outer envelope it will have a
spherical shell-like distribution while Fischer-Tropsch and
cometary hypotheses will produce a solid sphere-like distribution
since both require water to be formed much closer to the star.
Herschel Space Observatory will be able to distinguish between
these two possibilities since its beam size will be much narrower
than those of SWAS and ODIN. (2) The absence of pure rotational
and ro-vibrational lines of water in the ISO far-IR spectrum
analyzed by \citet{cer96iso} discards the presence of a
significant amount of water in the inner envelope. (3) As noted by
\citet{rod02}, the cometary origin implies the presence of HDO
with an abundance $\sim$0.06 \% relative to H$_2$O. HDO detection
would demonstrate that water is not formed from material ejected
by the star and the cometary hypothesis would be the only one
reasonable.\\

-- \textbf{HCO$^+$}. Its presence in IRC+10216 has been debated
for years. Chemical models \citep{gla86} predicted it with an
abundance high enough to allow its detection with available
millimeter or submillimeter telescopes. On the other hand,
observations did not clearly probe its existence. The $J$=1-0 line
(with an intensity of $T_A^*$=20 mK with the IRAM 30m telescope,
see Fig.~\ref{fig-spec-other}) often has been observed as a weak
feature at the noise level \citep{luc90}. The $J$=4-3 line was
observed as a weak feature of $T_A^*$$\sim$25 mK with the JCMT
telescope by \citet{ave94}. However, they failed to detect the
$J$=3-2 line at a noise level of $\sim$10 mK, which was predicted
with $T_A^*$$\sim$45 mK on the basis of a rotational temperature
of 17 K obtained from the intensities of the $J$=1-0 and $J$=4-3
lines. These considerations led these authors to conclude that the
feature at the $J$=4-3 frequency was not arising from HCO$^+$ and
they calculated an upper limit for its column density of
1.4$\times$10$^{11}$ cm$^{-2}$. Our Monte Carlo radiative transfer
calculations show that the excitation temperatures of each line
are significantly different, thus a prediction of the expected
intensities of other lines with a unique $T_{rot}$ should be taken
with care. To calculate the line profiles of HCO$^+$ we have
decreased by a factor of 2 the abundance profile obtained with the
chemical model in order to match the observed $J$=1-0 line.
HCO$^+$ is formed through the sequence of reactions R44+R28+R29
and therefore its abundance depends basically on the rates of
these reactions as well as on the destruction rates. The most
uncertain of all of them is the ionization by cosmic rays of H$_2$
(R44), which depends on the cosmic rays field in the surroundings
of IRC+10216. The abundance of HCO$^+$ approximately scales with
R44 rate. Therefore, a decrease by a factor of 2 in the cosmic
rays field would explain the decrease in abundance needed to match
the $J$=1-0 observed line.

The predictions for the first four pure rotational transitions are
shown in Fig.~\ref{fig-spec-other}. It is seen that the excitation
conditions are such that the $J$=3-2 line is expected with a
slightly higher intensity than the $J$=4-3 line, both being
$\sim$10 mK in $T_A^*$ for the JCMT telescope. This suggests that
the 25 mK feature observed by \citet{ave94} at the frequency of
$J$=4-3 is indeed not arising from HCO$^+$ because the $J$=3-2 was
not detected above a 10 mK noise level. Future long
time-integrated observations of these two lines will definitively
establish the excitations conditions of HCO$^+$ in IRC+10216.\\

-- \textbf{SO}. The abundance reached by SO in the chemical model
is enough to produce some lines with intensities of a few mK.
Fig.~\ref{fig-spec-other} shows the $N_J$=2$_3$-1$_2$ and
3$_4$-2$_3$ expected line intensities together with a feature at
99.300 GHz of a spectrum, obtained with the IRAM 30m telescope
\citep{cer06}, which could correspond to the 2$_3$-1$_2$
transition. A higher signal-to-noise ratio for this spectrum is
needed to claim detection. Long time-integrated observations at
the frequencies of other SO transitions (e.g. the 3$_4$-2$_3$)
would decide about the existence of SO in IRC+10216 formed trough
the ion-molecule chemical sequence of R32+R33
followed by the neutral-neutral reaction R20.\\

-- \textbf{H$_2$CS}. The chemical model predicts that this
molecule is produced with a moderate abundance, both through
neutral chemistry (from S+CH$_3$, R23) and through ionic chemistry
(from H$_3$CS$^+$, R39). The abundance reached is within the same
order of magnitude than for H$_2$CO, but less intense rotational
lines are expected due to its smaller dipole moment (1.649 D
versus 2.331 D of formaldehyde). We have also assumed an
ortho-to-para ratio of 3:1. Fig.~\ref{fig-spec-other} shows the
expected line profiles of ortho and para H$_2$CS, together with
some features from a $\lambda$ 3 mm spectrum taken with the IRAM
30m telescope \citep{cer06}. The features have intensities
somewhat higher than the predicted ones, but with the right order
of magnitude.

\section{Conclusions}

LTE chemical models predict that O--bearing species such as water
are very abundant in the inner envelope of IRC+10216. However, a
more realistic non-LTE approach, based on chemical kinetics,
indicates that the transformation of CO into H$_2$O within the gas
phase is not efficient in the inner layers, because of the high
energies required to break the CO bond. On the other hand, the
increase of SiO abundance with radius, predicted in LTE, is
possible because the reaction Si+CO$\rightarrow$SiO+C is
competitive in a high density scenario for the inner envelope. An
alternative mechanism for water production out of the gas phase
could be related to grain surfaces, which can act as a catalyst
reducing the activation energy for either a CO$\rightarrow$H$_2$O
process (due to Fischer-Tropsch catalysis according to
\citealt{wil04}) or a SiO$\rightarrow$H$_2$O process (on SiC
grains).

Concerning oxygen chemistry in the outer envelope, the release of
atomic oxygen due to CO photodissociation allows for the formation
of some O--bearing species. The abundance predicted for H$_2$CO is
a factor $\sim$5 lower than the observational estimation of
\citet{for04}. Wether formaldehyde is formed by the reaction
O+CH$_3$$\rightarrow$H$_2$CO+H or it is produced from
photodissociation of an unknown parent species released from
comets, as suggested by \citet{for04}, the expected shape of the
$\lambda$ 1.3 mm rotational lines disagrees with the observations.
An alternative source of H$_2$CO is suggested to be the radiative
association HCO$^+$+H$_2$$\rightarrow$H$_3$CO$^+$+h$\nu$. The
possibility of forming water in the outer envelope depends on the
rate constant of the radiative association between atomic oxygen
and molecular hydrogen. Non-local radiative transfer models show
that a rate constant as high as 10$^{-15}$ cm$^3$ s$^{-1}$ is
needed to reproduce the 556.936 GHz line profile observed by SWAS
and ODIN telescopes. Quantum chemical calculations of the rate
constants of the two radiative associations suggested in this
paper, specially the H$_2$+O reaction, could support or discard
them as important reactions in astrochemistry.

Other oxygen bearing species such as SO could exist with
abundances and excitation conditions which would produce
rotational lines near the detection limit of IRAM-30m telescope.
HCO$^+$ is observed with an intensity roughly in agreement with
that predicted from chemical and radiative transfer models.
Thioformaldehyde formation is also predicted through a chemical
route analogous to that of formaldehyde. This is supported by the
agreement of the line profiles obtained from radiative transfer
models with observations at $\lambda$ 3 mm.

Although the evaporation of cometary ices could be the source of
water in this late--stage carbon star, other phenomena such as
catalysis on grains or the production in the outer envelope
through the radiative association of H$_2$+O could also explain
the formation of this O--bearing species in the expanding
carbon-rich envelope. Future observations with the Herschel Space
Observatory will permit to distinguish the most plausible chemical
processes leading to the formation of water vapor in a carbon-rich
environment.

\acknowledgments

We thank J. R. Pardo for critical reading of the paper and very
useful suggestions and D. A. Neufeld for reading of the paper and
advice on water chemistry in diffuse clouds. We also acknowledge
T. J. Millar, E. Herbst and I. W. M. Smith, who kindly answered to
questions about basic concepts of chemistry during the completion
of this article. This work has been supported by Spanish MEC
grants AYA2003-2785 and ESP2004-665, and by Spanish ASTROCAM
S-0505/ESP-0237. MA also acknowledges funding support from Spanish
MEC through grant AP2003-4619.

\appendix

\section{Nucleation and Growth of SiC grains}

Here we describe the formation of SiC grains in the inner envelope
of IRC+10216 as a two-step process: (1) formation of condensation
nuclei plus (2) growth of grains from these nuclei by accretion of
gas phase species.

(1) The mathematical description of the nucleation process is
mainly taken from \citet{gai84}. We assume that the expanding gas
has at all moments a population of clusters (SiC)$_N$, composed of
different number $N$ of SiC monomers, which in thermal equilibrium
is given by
\begin{equation}
\displaystyle n_{(SiC)_N} = n_{SiC} \times exp\big(-\Delta
G(N)/kT_{gr}\big)
\end{equation}
where $n$ means numerical density, $T_{gr}$ is the grain
temperature and $\Delta G(N)$ is the free energy of formation of a
cluster of size $N$ from the monomers. The temperature of SiC
grains $T_{gr}$ is not equal to the kinetic temperature of the gas
$T_k$ because the former is affected by an inverse greenhouse
effect which makes $T_{gr}$ to be less than $T_k$ \citep{mcc82}.
To calculate $T_{gr}$ we follow the treatment of this author. The
magnitude $\Delta G(N)$ can be expressed as
\begin{equation}
\Delta G(N) = k \theta_N(T_{gr}) (N-1)^{2/3} - k T_{gr} (N-1) ln S
\end{equation}
where $S$ is the supersaturation ratio (ratio of the SiC vapor
pressure to its saturation pressure) that is calculated following
\citet{mcc82}, and the magnitude $\theta_{N}(T_{gr})$ (described
in \citealt{dra79}) is related to the surface tension of solid SiC
for large $N$ while for low $N$ is treated as a free parameter
since it is unknown.

Once we have establish a population of clusters (SiC)$_N$ given by
thermal equlibrium, nucleation theory says that there exist one
critical size $N_*$. Cluster with sizes $N$$\geq$$N_*$ will
continue accreting growth species (here SiC molecules) resulting
in grains of growing size, while clusters with sizes $N$$<$$N_*$
will revert to the monomers. Thus, there exists a bottleneck at
$N_*$ in the size spectrum. The critical size $N_*$ is found by
maximization of $\Delta G(N)$ with respect to $N$. The steady
state rate $J_*^S$ for the transformation
'(SiC)$_{N_*}$$\rightarrow$nucleus' is given by
\begin{equation}
\displaystyle J_*^S = \underbrace{\bigg\{\sum_{i=1}^{N_0} n_i
\big(\frac{k T_k}{2 \pi m_i}\big)^{1/2} \alpha_i\bigg\}}_{\beta}
\times \underbrace{4 \pi a_0^2 N_*^{2/3}}_{A_{N_*}} \times
n_{(SiC)_{N_*}} \times Z
\end{equation}
The subscript $i$ refers to the growth species: $n_i$ is the
numerical density, $m_i$ the mass, $\alpha_i$ the sticking
coefficient, and the sum of all contributions $\beta$ has the
meaning of a flux of particles. The subscript $*$ refers to
critical clusters: $A_{N_*}$ is the surface area and $a_0$ is the
radius of a monomer in the SiC lattice. $Z$ is the Zeldovich
factor (see Section 2 of \citealt{gai84} for a detailed
explanation).

Once $J_*^S$ is computed for the conditions prevailing at each
radius, the temporal evolution of the nucleation rate $J_*$ is
obtained from $J_*^S$ just considering a time lag $\tau_*$ (see
\citealt{gai84})
\begin{equation}
\displaystyle \frac{\partial}{\partial t} \Big\{
\frac{J_{*}}{\beta A_{N_{*}}} \Big\} = -\frac{1}{\tau_{*}} \Big\{
\frac{J_{*} - J_{*}^{S}}{\beta A_{N_{*}}} \Big\}
\end{equation}

(2) The growth of grains occur by addition of species that
incorporate into the grain, resulting in a temporal evolution for
the grain radius $a_{gr}$ of
\begin{equation}
\displaystyle \frac{\partial a_{gr}}{\partial t} =
\sum_{i}^{N_{0}} \alpha_{i} \times \Big(\frac{k T_k}{2 \pi
m_{i}}\Big)^{1/2} \times (4 \pi a_{gr}^{2}) \times n_{i} \times
\frac{a_{0,i}}{3 N_{i}^{2/3}}
\end{equation}
where the sum extends to all gaseous species $i$ that produces
growth of grains (here only SiC molecules are considered). The
parameters $\alpha_i$ and $m_i$ have the same meaning as in
equation A3, $a_{0,i}$ is the radius of the species $i$ in the
lattice and $N_i$ is the number of monomers of type $i$ forming
part of a grain.

The unknown parameters, such as the sticking coefficients and the
form of the function $\theta_N$ for low $N$ are chosen for
obtaining an abundance and size of SiC grains in agreement with
the literature: abundance relative to H$_2$ of $\sim$10$^{-14}$
and radius of $\sim$0.1 $\mu$m \citep{kna85,lor93}.

\section{Visual Extinction Radial Profile}

The parameters $A_i$ and $C_i$ in equation 8 are usually given in
the literature for plane-parallel geometry. Therefore, the
evaluation of the photodissociation rate $\Gamma_i$ of a species
$i$ through equation 8 is only valid for a species situated at
point P in a plane-parallel layer (see Fig.~\ref{fig-fis-out-av}
\notetoeditor{Fig.~\ref{fig-fis-out-av} should appear in the
appendix B section}), where the visual extinction measured along
the direction perpendicular to the infinite plane is $A_v^{pp}$.
But we want to know $\Gamma_i$ at the different points $r$ in
spherical geometry. The strategy, then, is to find expressions for
the UV field ($4\pi$$J_{UV}$) at points $P$ and $r$ in
plane-parallel and spherical geometries respectively. Equaling
both expressions we will determine the associated A$_v^{pp}$, that
we will insert in equation 8.

\begin{figure}[h]
\begin{center}
\includegraphics[angle=0,scale=.47]{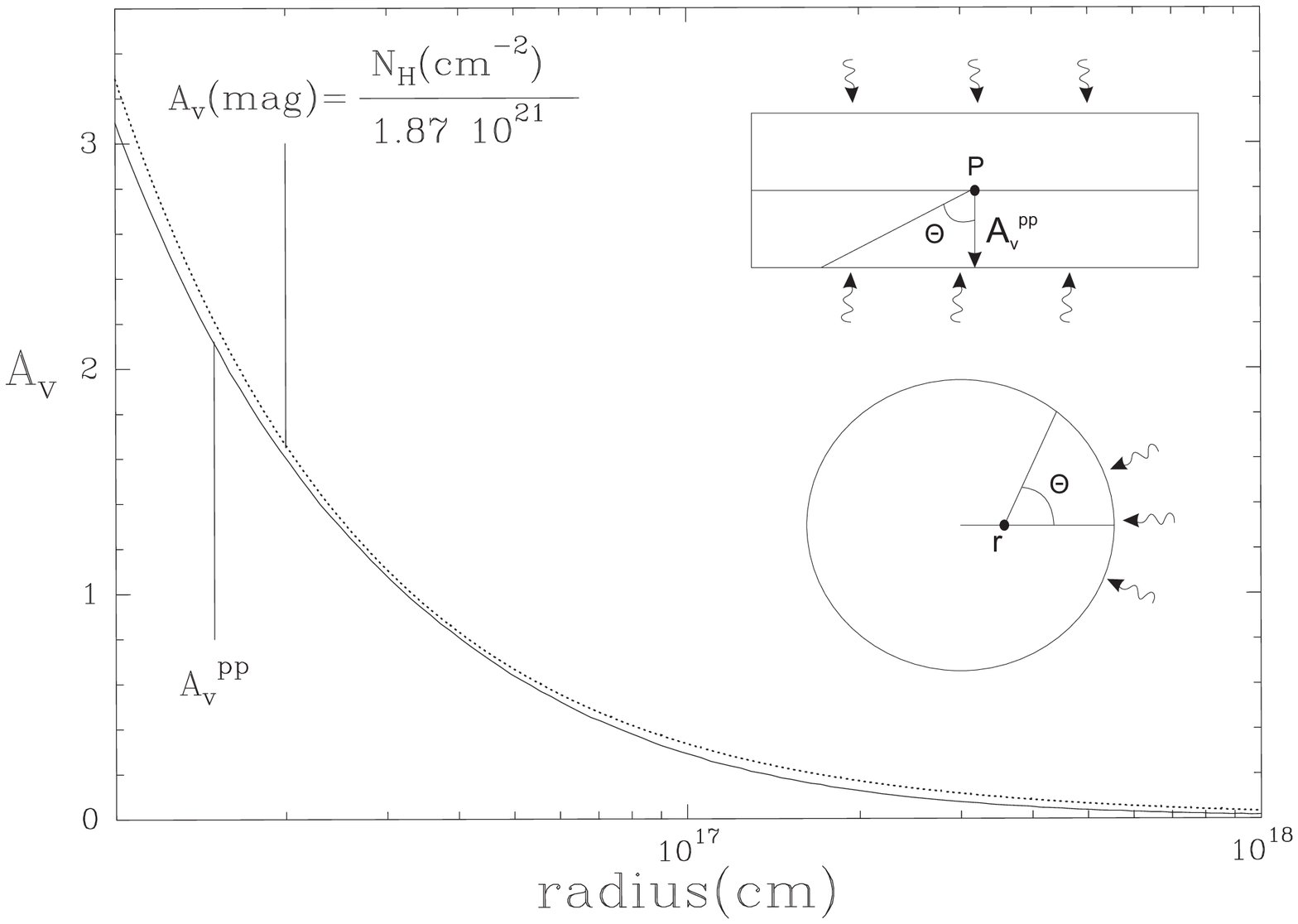}
\caption{$A_v$ radial profiles obtained from the plane-parallel to
spherical correction ($A_v^{pp}$) and from the standard relation
$A_v$$\varpropto$$N_H$ of \citet{boh78}. The inserts show a two
dimensions scheme of plane-parallel and spherical geometries.}
\label{fig-fis-out-av}
\end{center}
\end{figure}

The UV field at point $P$ in a constant density plane-parallel
layer is given by
\begin{equation}
4\pi J_{UV}(P) = 2\pi G_{0} \int_{0}^{\pi} sin\theta
exp\big\{-\tau_{1000}(\theta)\big\} d\theta  \quad , \quad
\bigg[\tau_{1000}(\theta) =
\Big(\frac{\tau_{1000}}{\tau_{5500}}\Big) \frac{A_{v}^{pp}}{1.086}
\frac{1}{cos\theta}\bigg]
\end{equation}
where $G_0$ is the wavelength integrated specific intensity of the
interstellar standard UV field \citep{dra78} (i.e. in photons
cm$^{-2}$ s$^{-1}$ sr$^{-1}$), $\theta$ is the angle of the
considered direction with respect to the normal to the surface.
The UV opacity $\tau_{1000}$ has been expressed as a function of
the visual extinction $A_v^{pp}$ through the ratio of
UV--to--visible opacities ($\tau_{1000}$/$\tau_{5500}$), where the
subscript stands for the mean wavelength in $\AA$.

The UV field at point $r$ within a sphere with a $r^{-2}$ density
law is \notetoeditor{Equation B2 might fit in one line in the
journal edition}
\begin{equation}
4\pi J_{UV}(r) = 2\pi G_{0} \int_{0}^{\pi} sin\theta
exp\big\{-\tau_{1000}(r,\theta)\big\} d\theta \quad , \quad
\bigg[\tau_{1000}(r,\theta)=\frac{\tau_{1000}^* r^*}{r \sqrt{1 +
cos^2\theta}} \Big\{\frac{\pi}{2} - atan
\big(\frac{-cos\theta}{\sqrt{1+cos^2\theta}}\big)\Big\}\bigg]
\end{equation}
where it is necessary a boundary condition, e.g. the UV optical
depth $\tau_{1000}^*$ at point $r^*$ along the radial direction.

We have assumed $\tau_{1000}$($r$=10$^{16}$)=12.7 \citep{dot98},
although the opacity at UV wavelengths is highly uncertain for
IRC+10216, and a ($\tau_{1000}/\tau_{5500}$) ratio of 3.1 as
measured by \citet{rou91} for amorphous carbon grains of 0.05
$\mu$m, although different values up to 8 can be found in the
literature depending on grain nature and size. The radial profile
$A_v(r)$ obtained is shown in Fig.~\ref{fig-fis-out-av} where it
is compared to the standard relation $A_v$$\varpropto$$N_H$ found
for the interstellar medium by \citet{boh78}. The interstellar UV
field $G$ has been decreased by a factor 2 with respect to the
standard $G_{0}$ to make the abundance distributions of C$_2$H and
CN (photodissociation products of C$_2$H$_2$ and HCN respectively)
to peak at $\sim$15'' and $\sim$20'' respectively in agreement
with interferometric observations \citep{day95,gue99}. For our
assumed distance of the star (150 pc) C$_2$H peaks at
3.4$\times$10$^{16}$ cm and CN at 4.5$\times$10$^{16}$ cm (see
Fig.~\ref{fig-quim-out}). We point out that this procedure is
affected by a degeneracy between at least two parameters: the star
distance, which transforms radial distances to angular distances,
and the UV field and/or the $A_v$ radial profile, which make
molecules photodissociate at a shorter or larger radial distance.
Despite the uncertainties in the knowledge of these quantities,
and even if our assumption of a less intense UV field than the
standard is not correct, the abundances of the different molecules
will be predicted at the right angular positions.

\end{document}